\documentclass[a4paper,twocolumn, iop, twocolappendix, numberedappendix]{oja}
\usepackage{graphicx} 
\usepackage[T1]{fontenc}
\usepackage[utf8]{inputenc}
\usepackage{natbib}
\usepackage{multirow}
\usepackage{comment}
\usepackage{graphicx}	
\usepackage{amssymb}	
\usepackage{xspace}        
\usepackage{xcolor}          
\usepackage{newtxtext,newtxmath}
\usepackage{booktabs}
\usepackage{fancyhdr}
\usepackage{upgreek}
\usepackage{times}

\usepackage[colorlinks,linkcolor=blue,citecolor=blue,urlcolor=blue ]{hyperref}
\usepackage{xurl} 
\usepackage{cleveref}
\usepackage{tikz}
\usetikzlibrary{shapes.geometric, arrows, calc}
\tikzset{every node/.style={inner sep=10pt,minimum height=1cm}}
\tikzstyle{startstop} = [rectangle, rounded corners, minimum width=1cm, minimum height=1cm,text centered, draw=black, fill=blue!50]
\tikzstyle{io} = [trapezium, trapezium left angle=90, trapezium right angle=90, minimum width=1cm, text width=5cm, minimum height=1cm, text centered, draw=black, fill=blue!30]
\tikzstyle{process} = [rectangle, minimum width=7cm, minimum height=1cm, text centered, text width=7cm, draw=black, fill=gray!20]
\tikzstyle{middle} = [trapezium, trapezium left angle=90, trapezium right angle=90, minimum width=0cm, text width=7cm, minimum height=1cm, text centered, draw=black, fill=green!10]
\tikzstyle{middle2} = [trapezium, trapezium left angle=80, trapezium right angle=100, minimum width=0cm, text width=7cm, minimum height=1cm, text centered, draw=black, fill=green!10]
\tikzstyle{subio} = [rectangle, minimum width=1cm, minimum height=1cm, text centered, draw=black, fill=blue!30]
\tikzstyle{arrow} = [thick,->,>=stealth]

\usepackage{cleveref}

\newcommand{\review}[1]{{\color{black}{#1}}}

\begin{document}
\title{Constraining the dispersion measure redshift relation with simulation-based inference}
\author{Koustav Konar$^\star$}

\affiliation{Ruhr University Bochum, Faculty of Physics and Astronomy, Astronomical Institute (AIRUB), German Centre for Cosmological
Lensing, 44780 Bochum, Germany}

\author{Robert Reischke$^{\dagger}$}
\affiliation{Argelander-Institut für Astronomie, Universität Bonn, Auf dem Hügel 71, D-53121 Bonn, Germany}
\affiliation{Ruhr University Bochum, Faculty of Physics and Astronomy, Astronomical Institute (AIRUB), German Centre for Cosmological
Lensing, 44780 Bochum, Germany}
\thanks{$^\dagger$\href{mailto:reischke@posteo.net}{reischke@posteo.net}}

\author{Steffen Hagstotz}
\affiliation{Universitäts-Sternwarte, Fakultät für Physik, Ludwig-Maximilians Universität München, 
Scheinerstraße 1, D-81679 München, Germany and\\
Excellence Cluster ORIGINS, Boltzmannstraße 2, D-85748 Garching, Germany}

\author{Andrina Nicola}
\affiliation{Argelander-Institut für Astronomie, Universität Bonn, Auf dem Hügel 71, D-53121 Bonn, Germany}

\author{Hendrik Hildebrandt}
\affiliation{Ruhr University Bochum, Faculty of Physics and Astronomy, Astronomical Institute (AIRUB), German Centre for Cosmological
Lensing, 44780 Bochum, Germany}

\begin{abstract}
We use the dispersion measure (DM) of localised Fast Radio Bursts (FRBs) to constrain cosmological and host galaxy parameters using simulation-based inference (SBI) for the first time. By simulating the large-scale structure of the electron density with the Generator for Large-Scale Structure (GLASS), we generate log-normal realisations of the free electron density field, accurately capturing the correlations between different FRBs.

For the host galaxy contribution, we rigorously test various models, including log-normal, truncated Gaussian and Gamma distributions, while modelling the Milky Way component using pulsar data. Through these simulations, we employ the truncated sequential neural posterior estimation method to obtain the posterior. Using current observational data, we successfully recover the amplitude of the DM-redshift relation, consistent with Planck, while also fitting both the mean host contribution and its shape. Notably, we find no clear preference for a specific model of the host galaxy contribution.

Although SBI may not yet be strictly necessary for FRB inference, this work lays the groundwork for the future, as the increasing volume of FRB data will demand precise modelling of both the host and large-scale structure components. Our modular simulation pipeline offers flexibility, allowing for easy integration of improved models as they become available, ensuring scalability and adaptability for upcoming analyses using FRBs. The pipeline is made publicly available under \href{https://github.com/koustav-konar/FastNeuralBurst}{https://github.com/koustav-konar/FastNeuralBurst}.
\end{abstract}

\maketitle
\section{Introduction}
Fast Radio Bursts (FRB) have received significant attention over the past decades, both from cosmological and astrophysical perspectives. First discovered in archival data \citep{lorimer_bright_2007}, these broad, millisecond transient pulses in the radio frequency range get dispersed by free electrons along their line of sight. While their origin is still debated \citep{petroff_fast_2019} and ranges from Magnetars \citep{thornton_population_2013,bochenek_fast_2020} to binary mergers \citep{liu2016fast}, it is clear that the majority of them must be of extragalactic origin due to their highly dispersed signal. The proportionality constant of this dispersion, fittingly called Dispersion Measure (DM), is proportional to the column density of electrons along the line-of-sight.

As a consequence, FRBs have been proposed to be used as a cosmological probe, in particular of the baryon distribution in the Universe. As for all cosmological fields, the electron density can be split into a background and fluctuation component relative to the background. If the host of the FRB is identified, an independent redshift estimate can be obtained. This allows the construction of the DM-$z$ relation, similar to the luminosity distance from supernovae. This relation has been used with current data to measure the baryon density and the Hubble constant \citep[e.g.][]{zhou_fast_2014,walters_future_2018,hagstotz_new_2022,macquart_census_2020,Beniamini_2021,wu_8_2022,james_measurement_2022,reischke_consistent_2023-1}. 
Likewise, one can study the statistical properties of the DM fluctuations \citep[e.g.][]{masui_dispersion_2015,shirasaki_large-scale_2017,rafiei-ravandi_chimefrb_2021,bhattacharya_fast_2020,takahashi_statistical_2021,reischke_probing_2021,  reischke_consistent_2022,reischke_calibrating_2023}. 

The ever-increasing number of observed FRBs \citep[currently around 600 unique events, see e.g.][]{newburgh_hirax_2016,chime_2021_first, khrykin_flimflam_2024} leads to raising interest in these events.
 The Square Kilometre Array \citep[SKA\footnote{\url{https://www.skao.int/}},][]{dewdney_2009_square} should observe $>\!10^5$  FRBs. Also, other surveys like DSA-2000 \citep{2019BAAS...51g.255H} are planning to detect $>\!10^4$ FRBs with host identification. This increasing number makes the modelling and inference process prone to systematic effects. \citet{reischke_cosmological_2023} showed that already with $\sim300 $ FRBs, it becomes necessary to include their covariance to conduct unbiased parameter inference using the DM-$z$ relation, an effect which has been neglected in all studies so far. With around $10^4$ FRBs, additional effects such as magnification can become important as well \citep{2024arXiv240706621T}. Using FRBs as a tool for cosmology and astrophysics, therefore, requires careful modelling. A lot of these effects can be challenging to model analytically, including the case where systematic effects from the search (or in cosmological terms, survey) strategy will not be tractable.

In this paper, we want to tackle these issues and present simulation-based inference (SBI) of cosmological and astrophysical models via the DM-$z$ relation of FRBs. SBI, sometimes also referred to as Likelihood-Free Inference or Implicit Likelihood Inference, is a Bayesian inference technique that does not require an explicit expression for the likelihood function of the data given the parameters of interest. Instead, the likelihood is implicitly assessed by evaluating the joint probability of the data and parameters from forward simulations that map the parameters to the corresponding synthetic data vectors. This approach offers several advantages over traditional methods that necessitate an explicit form for the likelihood. Firstly, the likelihood can assume any form, thus allowing one to bypass the common assumption of a Gaussian likelihood or the need to define a complex analytical expression for the likelihood. Secondly, for certain models and measurements, it might be impractical or too resource-intensive to determine an analytical likelihood.
On the similar side, for SBI, data compression becomes essential for the high-dimensional data and parameter spaces typical in cosmology \citep{leclercq2018bayesian,alsing_fast_2019}. \review{Although not strictly necessary, integrating data compression within the SBI framework can reduce the network training time as the resulting data vector is low-dimensional. Additionally, it reduces the number of simulations required for the inference by effectively reducing the size of the data vector. 
A downside can be, however, that data compression is not always completely lossless.} The inference methods available in the SBI framework also vary based on their complexities, from the relatively trivial Approximate Bayesian Computing \citep[ABC, see e.g.][]{rubin1984bayesianly,pritchard1999population,beaumont2019approximate} to the latest development in neural network (NN). \review{Both procedures, SBI and data compression, have been applied to cosmological data analysis with the help of NNs. For example, \citet{2023MNRAS.521.2050L,2023A&A...675A.120E} use neural compression to reduce the dimensionality of the data space but follow traditional likelihood analysis on the (NN-estimated) summary statistics, with \citet{2022PhRvD.105h3518F} also using ABC on top of this compression. In contrast, \citet{2023MNRAS.524.6167L,von_wietersheim-kramsta_kids-sbi_2024,2024arXiv240510881G} indeed use SBI with neutral density estimation (NDE) but simpler compression techniques such as score compression.}

NDE requires forward simulations to learn the posterior distribution $p(\boldsymbol{\theta}|\boldsymbol{d})$ where $\boldsymbol{d}$ is simulated given $\boldsymbol{\theta}$. In our case, these forward simulations consist in principle of three components: $(i)$ The large-scale structure (LSS) and the background component are produced using the Generator for Large-Scale Structure \citep[GLASS,][]{tessore_glass_2023} using halo model power spectra for the three-dimensional electron power spectrum. This will generate log-normal realisations of the electron field with the correct two-point statistics imprinted. $(ii)$ The host contribution, which is simply sampled from a host model probability density function (PDF). $(iii)$ The Milky Way (MW) contribution, for which we will use the standard methods of inferring it from already present electron models \citep{cordes2002ne2001,yao_new_2017,yamasaki_galactic_2020}.
We will then use those forward simulations to train a NN to learn the posterior and sample from the posterior with traditional MCMC.
Here, we will use Truncated Sequential Neural Posterior Estimation \citep[TSNPE,][]{deistler2022truncated} to conduct the inference within the SBI framework. We aim to fit the amplitude of the DM-$z$ relation and the median and the width of the log-normal host distribution with the available host-identified FRBs, providing a roadmap for future cosmological and astrophysical inference with FRBs.

The manuscript is structured as follows: In Section \ref{sec:dm}, we introduce the basics of FRB cosmology and discuss the different components entering the total DM. Section \ref{sec:simulated_dm} provides an overview of the forward simulation pipeline. The inference techniques are discussed in Section \ref{sec:inference}. Section \ref{sec:validation} introduces the validation techniques we use after the inference process. In Section \ref{sec:results}, we present the results and summarise them in Section \ref{sec:conclusion}.

\section{Dispersion Measure Components}
\label{sec:dm}
\subsection{FRB basics}
The pulses of FRBs undergo dispersion while travelling through the ionized matter distribution in the Universe, leading to a frequency-dependent, $\propto \nu^{-2}$, offset of the bursts' arrival times \citep[see e.g.][]{2007Sci...318..777L,2019A&ARv..27....4P}. Given this time delay measured as $\delta t(\hat{\boldsymbol{x}},z)$ for an FRB at redshift $z$ in direction $\hat{\boldsymbol{x}}$, the constant of proportionality is the observed dispersion measure: $\delta t(\hat{\boldsymbol{x}},z) = \mathrm{DM}_\mathrm{tot}(\hat{\boldsymbol{x}}, z) \nu^{-2}$.
This DM can be broken up into different components:
\begin{align}
\label{eq:dispersion_measure_contributions}
    \mathrm{DM}_\mathrm{tot}(\hat{\boldsymbol{x}},z) = \mathrm{DM}_\mathrm{LSS}(\hat{\boldsymbol{x}},z) + \mathrm{DM}_\mathrm{MW}(\hat{\boldsymbol{x}}) + \mathrm{DM}_\mathrm{host}(z)\;.
\end{align}
The first contribution is $\mathrm{DM}_\mathrm{LSS}(\hat{\boldsymbol{x}},z)$, caused by free electrons in the LSS. Here, the dependence on the direction is kept explicitly since the LSS is correlated. In the literature, $\mathrm{DM}_\mathrm{LSS}(\hat{\boldsymbol{x}},z)$ is often split up into an IGM part and a halo part
\begin{align}
    \mathrm{DM}_\mathrm{LSS}(\hat{\boldsymbol{x}},z) = \mathrm{DM}_\mathrm{IGM}(\hat{\boldsymbol{x}},z) + \mathrm{DM}_\mathrm{halo}(\hat{\boldsymbol{x}},z)\;.
\end{align}
This is equivalent to a halo model prescription \citep{cooray_halo_2002} of the statistical properties of the DM. On the level of the power spectrum, this would amount to the two-halo term (corresponding to $\mathrm{DM}_\mathrm{IGM}(\hat{\boldsymbol{x}},z)$) and the one-halo term (corresponding to $\mathrm{DM}_\mathrm{halo}(\hat{\boldsymbol{x}},z)$).

The MW contribution $\mathrm{DM}_\mathrm{MW}(\hat{\boldsymbol{x}})$ can itself be split up into a contribution from the ISM and the MW halo. Both will not depend on redshift, as these are local quantities. However, there is a clear directional dependence. Lastly, $\mathrm{DM}_\mathrm{host}(z)$ is the contribution of the host galaxy which can, as the MW contribution, be split up into a part originating from the visible galaxy and one of the halo. For this, only a potential redshift dependence is assumed, as the contribution of different hosts should not be correlated, ignoring the unlikely event that two distinct FRBs originate from the same galaxy.
Note that the rest-frame DM of the host, $\mathrm{DM}_\mathrm{host,rf}$, is observed as $\mathrm{DM}_\mathrm{host}(z) = (1+z)^{-1}\mathrm{DM}_\mathrm{host,rf}$. 

\subsection{Large-Scale-Structure Contribution}
First, we will take a more detailed look at the contribution of the LSS. Quite generally, this is given by
\begin{align}
\label{eq:DM_LSS}
    \mathrm{DM}_\mathrm{LSS}(\hat{\boldsymbol
    {x}},z) = \int_0^z \! n_\mathrm{e}(\hat{\boldsymbol
    {x}},z') \, f_\mathrm{IGM}(z') \, \frac{1+z'}{H(z')} \, \mathrm d z' \; ,
\end{align}
where $n_\mathrm{e}(\hat{\boldsymbol{x}},z)$ is the {comoving} cosmic free electron density, $H(z) = H_0 E(z)$ is the Hubble function with the expansion function $E(z)$ and the Hubble constant $H_0$. $f_\mathrm{IGM}(z)$ is the fraction of electrons in the IGM and is calculated by subtracting the fraction bound in stars, compact objects and the dense interstellar medium (ISM)
\begin{align}
\label{eq:f_IGM}
    f_\mathrm{IGM}(z) = 1 - f_\star(z) - f_\mathrm{ISM}(z) \, .
\end{align}
For redshifts $z<3$, almost all baryons are ionised, and the DM is, therefore, rewritten as
\begin{align}
\label{eq:n_e}
    n_\mathrm{e}(\hat{\boldsymbol
    {x}},z) =  \chi_\mathrm{e} \frac{\rho_\mathrm{b}(\hat{\boldsymbol
    {x}}, z)}{m_\mathrm{p}} = \chi_\mathrm{e} \frac{\bar \rho_\mathrm{b}}{m_\mathrm{p}} \big( 1 + \delta_\mathrm{e} (\hat{\boldsymbol
    {x}}, z) )\;,
\end{align}
with the baryon density $\rho_\mathrm{b}$, the proton mass $m_\mathrm{p}$ and the electron fraction
\begin{align}
\label{eq:chi_e}
    \chi_\mathrm{e} = Y_\mathrm{H} + \frac{1}{2} Y_\mathrm{He} 
     \approx 1 - \frac{1}{2} {Y}_\mathrm{He} \; ,
\end{align}
calculated from the primordial hydrogen and helium abundances $Y_\mathrm{H}$ and $Y_\mathrm{He}$. Altogether, one finds:
\begin{align}
\label{eq:DM_LSS_v2}
    \mathrm{DM}_\mathrm{LSS}(\hat{\boldsymbol{x}},z) = \mathcal{A} \int_0^z \, \frac{1+z'}{E(z')}  \big(1+\delta_\mathrm{e}(\hat{\boldsymbol{x}},z')\big) \mathrm{d} z' \; ,
\end{align}
where we defined $\mathcal{A}\coloneqq \frac{3 c \Omega_\mathrm{b0} H_0}{8 \pi G m_\mathrm{p}} \chi_\mathrm{e} \, f_\mathrm{IGM}$. 
The LSS contribution is therefore entirely specified by the statistical properties of the electron density field $\delta_\mathrm{e}$.
Motivated by numerical simulations of the DM which showed that it follows a log-normal distribution \citep[see e.g.][]{zhang_intergalactic_2021,2024A&A...683A..71W}, we model the electron field using \texttt{GLASS} \citep{tessore_glass_2023} which can, given a three-dimensional power spectrum of a cosmological field, create log-normal realisations. This is done by dividing the LSS into $N_\mathrm{shells}$ non-overlapping and concentric shells. Each shell covers the full sky, spanning the comoving volume between redshift $z_i$ and $z_{i+1}$. First, we define a matter weight function along the line-of-sight via:
 \begin{align}
W^{(i)}(z) \coloneqq
\left\{
\begin{array}{cl}
   \chi^2(z)/E(z)  &\mathrm{if} \; z_i \leq z < z_{i+1}\;,  \\
    0  & \mathrm{else}\;,
\end{array}\right.
 \end{align}
with the co-moving distance $\chi(z)$. The electron density contrast, $\delta_\mathrm{e}(\hat{\boldsymbol{x}},z')$, can now also be defined in each shell:
\begin{equation}
\label{eq:shells}
   \delta^{(i)}_\mathrm{e} (\hat{\boldsymbol{x}}) = \int\mathrm{d}z\; W^{(i)}(z)\delta_\mathrm{e}(\hat{\boldsymbol{x}},z)\;.
\end{equation}
The statistical properties of this field on the two-point level are calculated in the harmonic space by the angular power spectra:
\begin{multline}
    C_{\delta_\mathrm{e} \delta_\mathrm{e}}^{(i j)}(\ell) = \frac{2}{\uppi} \int \mathrm{d}\chi \, W^{(i)}(z(\chi)) \int \mathrm{d}\chi' \, W^{(j)}(z(\chi^\prime)) \\ \int \mathrm{d}k \, k^{2} \,  P_{\delta_\mathrm{e}, \mathrm{nl}}(k, z(\chi), z(\chi^\prime)) \, j_{\ell}(k \chi) \, j_{\ell}(k \chi^\prime)\;,
    \label{eq:method:fs:nonlimber:non_limber}
\end{multline}
where $j_{\ell}(x)$ are spherical Bessel functions of order $\ell$ and $\ell \in \mathbb{N}$. The electron power spectrum is defined as:
 \begin{align}
     \langle\delta_\mathrm{e}(\boldsymbol{k},z) \delta_\mathrm{e}(\boldsymbol{k}^\prime,z^\prime)\rangle = (2\uppi)^3\delta_\mathrm{D}(\boldsymbol{k} + \boldsymbol{k}^\prime) P_{\delta_\mathrm{e}, \mathrm{nl}}(k, z, z^\prime)\;.
 \end{align}
 Lastly, we approximate the unequal time correlator by its geometric mean:
 \begin{equation}
     P_{\delta_\mathrm{e}, \mathrm{nl}}(k, z, z^\prime) \approx \left[P_{\delta_\mathrm{e}, \mathrm{nl}}(k, z)P_{\delta_\mathrm{e}, \mathrm{nl}}(k, z^\prime)\right]^{1/2}\,,
 \end{equation}
which has been shown to be an excellent approximation for weak gravitational lensing \citep{kitching_unequal-time_2017,de_la_bella_unequal-time_2021}. Since the DM of the LSS has a similarly broad kernel as it is an integrated effect, this should also hold for FRBs.
Starting from \Cref{eq:DM_LSS_v2}, let us define the perturbations to the DM as
\begin{equation}
    \mathcal{D}(\hat{\boldsymbol{x}},z) = \mathcal{A} \int_0^z \, \frac{1+z'}{E(z')}\delta_\mathrm{e}(\hat{\boldsymbol{x}},z') \mathrm{d} z' \;.
\end{equation}
It can be broken down into a discrete sum over the matter shells
\begin{equation}
\label{eq:method:DM_from_shell}
    \mathcal{D}(\hat{\boldsymbol{x}},z_N) \approx \mathcal{A}\sum_{i = 1}^{N_\mathrm{shells}}\frac{1+\bar{z}_i}{E(\bar{z}_i)} w_i\delta_{\mathrm{e},i}(\hat{\boldsymbol{x}})\;,
\end{equation}
where we defined the characteristic redshift of each shell to be its mean:
\begin{equation}
    \bar{{z}}^{(i)} =  \frac{\int \mathrm{d}{z} \, {z} \, {W}^{(i)}({z})}{\int \mathrm{d}{z}  \, {W}^{(i)}({z})}\;.
\end{equation}
Likewise, $w_i$ takes into account the weight of the shell via
\begin{equation}
\label{eq:method:weight_of_shells}
    w_i = \frac{1}{W^{(i)}(\bar{z}_i)}\int W^{(i)}(z)\mathrm{d}z\;.
\end{equation}

\subsection{Host Contribution}
\label{fig:host_contribution}
The next component in the forward simulation is the host contribution, quantifying the effect of the host galaxy on the observed DM. Based on the position of the FRB progenitor in the galaxy, the signal may travel through the whole galaxy or parts of it. As such, the induced DM varies accordingly. The effect of a complete or partial travel path through the local host, which translates to high and low DM, is usually described by a log-normal distribution \citep{macquart_census_2020, wu_8_2022}. Other works, e.g. \citet{hagstotz_new_2022}, have assumed a Gaussian host contribution. \review{Recently, simulations have also shown that a log-normal distribution can fit the host contribution rather well \citep{theis_galaxy_2024}. While the latter study made a quantitative comparison, as the statistical scatter in the simulations is difficult to estimate, a log-normal distribution provides a good fit. Furthermore, with the number of FRBs currently available, this model is accurate enough to capture the complexity of the data. Lastly, the log-normality can also be supported by analytical arguments given, for example, in \citet{mcquinn_locating_2014} and \citet{2024arXiv241117682R}.} With the current data set of FRBs, this choice does not make a difference if priors on the DM are included, as we will show later. For our fiducial case, however, we will choose a log-normal distribution for $\mathrm{DM}_\mathrm{host,rf}$ (that is the host contribution in the rest-frame of the host galaxy for which we assume no intrinsic redshift evolution):
\begin{align}
\label{eq:host:lognormal_pdf}
    p_\mathrm{host}(x; \mu, \sigma_\mathrm{LN}) = \frac{1}{ x \sigma_\mathrm{LN} \sqrt{2\pi}} \mathrm{exp} \left(- \frac{(\mathrm{ln}x - \mu)^2}{2\sigma_\mathrm{LN}^2} \right)\,,
\end{align}
{where $\mathrm{exp}(\mu)$ and $\mathrm{exp}(2\mu + \sigma_\mathrm{LN}^2)[\mathrm{exp}(\sigma_\mathrm{LN}^2) - 1]$ are the median and variance respectively with $x =\mathrm{DM}_\mathrm{host,rf}$.} The median and the scale ($\sigma_\mathrm{LN}$) are free parameters in our study. Note that this samples the rest-frame DM of the host, $\mathrm{DM}_\mathrm{host,rf}$. \review{Here we assume that the rest-frame host contribution is constant, although there are some hints for redshift evolution \citep[see e.g.][]{Acharya_2025}}.

\subsection{Milky Way contribution}
\label{sec:mw_contribution}

In our analysis, we assume that the MW contribution is accessible from the models described in \citet{cordes2002ne2001,yao_new_2017,yamasaki_galactic_2020} and we simply add the numerical value for $\mathrm{DM}_\mathrm{MW}(\hat{\boldsymbol{x}})$ in \Cref{eq:dispersion_measure_contributions}. At the current sensitivity level, dictated by the amounts of FRBs available, this addition does not have a sizable influence on the inference of cosmological parameters. However, with more FRBs being observed, it could well be that the addition of the MW contribution leads to a residual correlation in the DM, which will be falsely picked up as a cosmological signal. This scenario can, in principle, be tested with our pipeline, as the MW contribution can simply be added to the simulated DM.

\section{Model generation}
\label{sec:simulated_dm}
\subsection{Forward simulation}
In this section, we describe how we use the previously described ingredients to construct forward simulations for the individual components of \Cref{eq:dispersion_measure_contributions}, which, by simple summation, yield a forward simulation prediction for a set of $ \mathrm{DM}_\mathrm{tot}(\hat{\boldsymbol{x}}_a,z_a)$, with $a=1,\dots, N_\mathrm{FRB}$. Here, $\hat{\boldsymbol{x}}_a$ and $z_a$ are the positions and redshifts of the FRBs from the FRB catalogue. \Cref{fig:flowchart} summarises the forward simulation pipeline, which goes through the following steps:
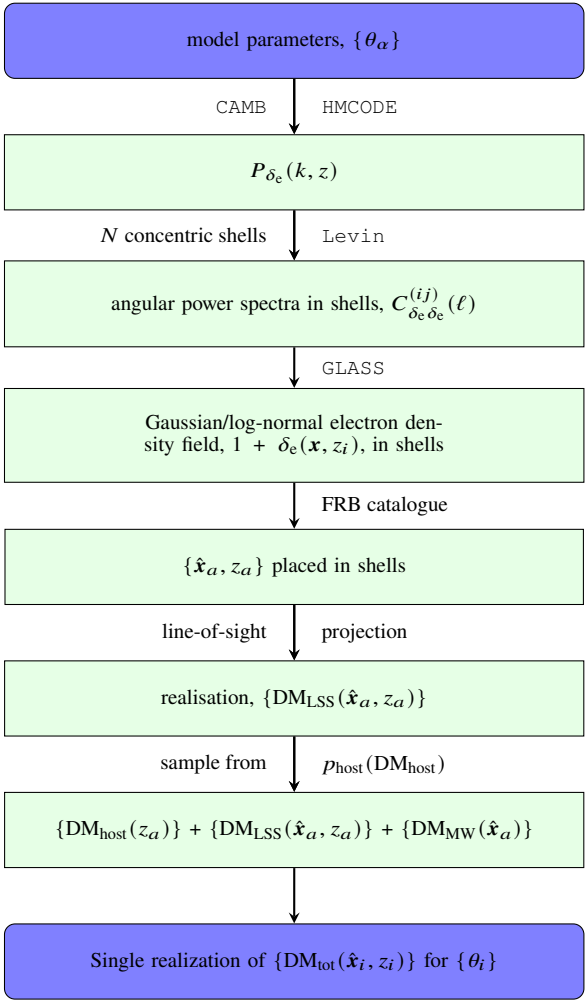
\begin{figure}
    \centering
    \begin{tikzpicture}[node distance=1.75cm]
        \node (start) [startstop, text width=7cm] {model parameters, $\{\theta_\alpha\}$};
        \node (camb) [middle, below of=start, text width=6.85cm] {$P_{\delta_\mathrm{e}}(k,z)$};
        \node (cl) [middle, below of=camb, text width=7cm] {angular power spectra in shells, $C_{\delta_\mathrm{e} \delta_\mathrm{e}}^{(i j)}(\ell)$};
         \node (random) [middle, below of=cl] {Gaussian/log-normal electron density field, $1 + \delta_\mathrm{e}(\boldsymbol{x},z_i)$, in shells};
        \node (glass) [middle, below of=random, text width=6.7cm] {$\{\hat{\boldsymbol{x}}_a,z_a\}$ placed in shells} ;
        \node (sampling) [middle, below of=glass, text width=6.67cm] {realisation, $\{\mathrm{DM}_\mathrm{LSS}(\hat{\boldsymbol{x}}_a,z_a)\}$};
        \node (adding) [middle, below of=sampling, text width=6.67cm] {$\{\mathrm{DM}_\mathrm{host}(z_a)\}$ + $\{\mathrm{DM}_\mathrm{LSS}(\hat{\boldsymbol{x}}_a,z_a)\}$ + $\{\mathrm{DM}_\mathrm{MW}(\hat{\boldsymbol{x}}_a)\}$};
        \node (dmtot) [startstop, below of=adding, text width=7cm] {Single realization of $\{\mathrm{DM}_\mathrm{tot}(\hat{\boldsymbol{x}}_i,z_i)\}$ for $\{\theta_i\}$};
        \draw [arrow] (start) -- node[anchor=east] {\texttt{CAMB}} (camb);
        \draw [arrow] (start) -- node[anchor=west] {\texttt{HMCODE}} (camb);
        \draw [arrow] (camb) -- node[anchor=east] {$N$ concentric shells} (cl);
        \draw [arrow] (camb) -- node[anchor=west] {\texttt{Levin}} (cl);
        \draw [arrow] (cl) -- node[anchor=west] {\texttt{GLASS}} (random);
        \draw [arrow] (random) -- node[anchor=west]{FRB catalogue}(glass);
         \draw [arrow] (glass) -- node[anchor=west]{projection}(sampling);
         \draw [arrow] (glass) -- node[anchor=east]{line-of-sight}(sampling);
         \draw [arrow] (sampling) --node[anchor=east]{sample from} (adding);
        \draw [arrow] (sampling) --node[anchor=west]{$p_\mathrm{host}(\mathrm{DM}_\mathrm{host})$} (adding);
        \draw [arrow] (adding) -- (dmtot);
    \end{tikzpicture}
\caption{Flowchart of the simulation pipeline described in Section \ref{sec:simulated_dm}. Blue boxes indicate the pipeline input and output. Green boxes show intermediate data products, and the labels of the arrows depict the operation applied to the previous data product.}
\label{fig:flowchart}
\end{figure}
\begin{enumerate}
    \item Fix the model parameters $\{\theta_\alpha\}$ and obtain the 3-dimensional non-linear electron power spectrum, $P_{\delta_\mathrm{e}\delta_\mathrm{e}}(k,z)$ using \texttt{CAMB} \citep{lewis_efficient_2000, lewis_cosmological_2002,howlett_cmb_2012} and then \texttt{HMCODE} \citep{mead_accurate_2015,mead_hydrodynamical_2020,tröster_2022_joint}.
    \item Define concentric shells such that there are no discreteness effects, i.e. that a finer resolution along the line-of-sight does not change the results. We found that $N_\mathrm{shells} = 17$ is enough for our purpose for $z\in[0.01,1]$. Use \texttt{Levin} \citep{zieser_cross-correlation_2016,leonard_n5k_2023} to calculate the angular power spectrum in those shells via \Cref{eq:method:fs:nonlimber:non_limber}.
    \item Run \texttt{GLASS} \citep{tessore_glass_2023} with the angular power spectra in the shells to generate log-normal or Gaussian realisations of the electron overdensity. Add unity to each shell to arrive at the physical density.
    \item Place all FRBs from the catalogue in the simulated electron density field and project it along the line of sight.
    If we label all shells $\mathrm{s}_i$ and $z<\mathrm{s}_i$ is interpreted as that $z$ is strictly below all redshifts in shell $i$, we can define an auxiliary weight as:
    \begin{align}
\label{eq:sims:weight_for_redshift_interpolation}
w^\mathrm{DM}_{i}(z_\mathrm{FRB}) \coloneqq
\left\{
\begin{array}{cl}
    0  &\mathrm{if} \; z_\mathrm{FRB} < \mathrm{s}_i \;, \\
    & \\ \;
    \frac{z_\mathrm{FRB} - z_\mathrm{i, min}}{z_\mathrm{i, max} - z_\mathrm{i,min}}  &\mathrm{if} \;  z_\mathrm{FRB} \in \mathrm{s}_i \;, \\
    & \\ \;
    1  &\mathrm{if} \; \mathrm{s}_i \leq z_\mathrm{FRB}  \;.
\end{array}\right.
\end{align}
Including this weight in \Cref{eq:method:DM_from_shell} one finds:
\begin{equation}
\label{eq:sims:DM_from_shell_with_redshift_interpolation}
    \resizebox{.71\hsize}{!}{$\mathcal{D}(\hat{\boldsymbol{x}}_a,z_a) = \mathcal{A}\sum_{i = 1}^{N}\frac{1+\bar{z}_i}{E(\bar{z}_i)} w_i^\mathrm{DM} (z_a)w_i\delta_{\mathrm{e},i}(\hat{\boldsymbol{x}}_a)\;.$}
\end{equation}
Lastly, adding the homogeneous contribution gives:
\begin{equation}
     \resizebox{.71\hsize}{!}{$\mathrm{DM}_\mathrm{LSS}(\hat{\boldsymbol{x}}_a,z_a) = \mathcal{A} \int_0^{z_a} \, \frac{1+z'}{E(z')}  \mathrm{d} z' + \mathcal{D}(\hat{\boldsymbol{x}}_a,z_a)  \; .$}
\end{equation}
    \item Draw samples from the host PDF contribution for each FRB (in our case \Cref{eq:host:lognormal_pdf}) and map it to the physical frame by redshifting it.
    \item {Obtain the MW contribution as discussed in Section \ref{sec:mw_contribution}.}
    \item Add all contributions together.
\end{enumerate}
This procedure provides a pair $(\{\theta_\alpha\},\{\mathrm{DM}_\mathrm{tot} (\hat{\boldsymbol{x}}_a,z_a)\})$. Rerunning the pipeline, with parameters sampled from a prior distribution, creates a set of forward simulations, which is used to learn the posterior distribution by the NN. {The resolution of the simulation is dependent on the parameter $N_\mathrm{side} \in 2^{\mathbb{Z}^+}$ as $\texttt{GLASS}$ internally uses $\texttt{HEALPix}$ \citep{gorski2005healpix}.} From the flowchart in \Cref{fig:flowchart} and the list above, it is clear that any component in the pipeline can easily be exchanged for another model. 
\review{If, for example, one wants to connect FRB observations closer to cosmological analysis and an effective model of the statistical properties of the gas distribution \citep[e.g.][]{schneider_new_2015,2024arXiv241117682R}, this can be easily done.} 

\subsection{Data compression}
\label{subsec:compression}
The data vector currently is $N_\mathrm{FRB}$ dimensional, thus requiring a compression procedure, essentially translating a $d$-dimensional dataset down to $n$ dimensions ($n$ < $d$). This is an essential step, as training the NN with high-dimensional data is slow and can lead to inaccuracies. Lossless data compression preserves the Fisher information of the original data.
Therefore, we use the data reduction scheme named score compression prescribed in \citet{alsing_fast_2019}, reducing the $d$-dimensional data down to a dimension equal to the number of free parameters. Assuming a Gaussian likelihood, this compressed data ($t$) can be obtained as \citep{tegmark_karhunen-loeve_1997}
\begin{equation}
\begin{split}
\label{eq:sims:data_compressed_data}
    \boldsymbol{t} = & \,\boldsymbol{\nabla\mu}_*^{\boldsymbol{T}} \boldsymbol{\boldsymbol{\mathrm{C}}}_*^{-1} (\boldsymbol{\mathrm{d}} - \boldsymbol{\mu}_*) \\ & + \frac{1}{2} (\boldsymbol{\mathrm{d}} - \boldsymbol{\mu}_*)^{\boldsymbol{T}} \boldsymbol{\boldsymbol{\mathrm{C}}}_*^{-1} \boldsymbol{\nabla\boldsymbol{\mathrm{C}}}_*\boldsymbol{\boldsymbol{\mathrm{C}}}_*^{-1}(\boldsymbol{\mathrm{d}} - \boldsymbol{\mu}_*)\;,    
\end{split}
\end{equation}
where $\boldsymbol{\mu}_*, \boldsymbol{\boldsymbol{\mathrm{C}}}_*$ are the ensemble mean and covariance of the original data at some fiducial value $\boldsymbol{\theta}_*$ and $\boldsymbol{\mathrm{d}}$ is the corresponding data vector. $\boldsymbol{\nabla}$ represents the partial derivatives with respect to the free parameters. \review{We note that the Gaussian likelihood assumption is only employed for the data compression and is removed after that. Hence, the inference itself remains likelihood-free without losing any generality. Improper likelihood assumption can, however, lead to suboptimal compression and this fiducial $\boldsymbol{\theta}_*$ needs to be optimised to ensure no information loss \citep{alsing2018massive}}. Specifically, we use the Fisher scoring method
\begin{align}
\label{eq:sims:data_fisher_scoring}
    \boldsymbol{\theta}_{k+1} = \boldsymbol{\theta}_k + \boldsymbol{\mathrm{F}}_k^{-1}\boldsymbol{t}_k\;,
\end{align}
where $\boldsymbol{{t}}_k$ is the compressed statistics for $\boldsymbol{\theta}_k$ at the ${k}$-th step. Depending on the complexity of the parameter space, a larger value of $k$ may be required for convergence. In practice, we stop after a finite number of iterations when the increment, the second term on the right-hand side of \Cref{eq:sims:data_fisher_scoring}, asymptotically approaches a plateau and further steps are not favoured against the computational time. The components of the Fisher matrix, $\mathrm{F}_{ij}$, are evaluated by its full expression for a Gaussian likelihood:
\begin{equation}
\begin{split}
\label{eq:sims:data_fisher_matrix}
    \mathrm{F}_{ij} = \, \frac{1}{2} &\mathrm{Tr}\Big[ \boldsymbol{\mathrm{C}}^{-1} \nabla_i\boldsymbol{\mathrm{C}} \ \boldsymbol{\mathrm{C}}^{-1}  \nabla_j\boldsymbol{\mathrm{C}} \\ &+\; \boldsymbol{\mathrm{C}}^{-1} (\nabla_i\boldsymbol{\mu} \nabla_j\boldsymbol{\mu}^{\mathrm{T}} + \nabla_i\boldsymbol{\mu}^{\mathrm{T}} \nabla_j\boldsymbol{\mu} ) \Big]\;.    
\end{split}
\end{equation}
The three free parameters in our model are the amplitude of the DM-$z$ relation, the median and the scale of the log-normal host distribution. We specify the initial $\boldsymbol{\theta}_*$ in \Cref{tab:parameters} motivated from \citet{reischke_consistent_2022}, but we quickly converge to $\boldsymbol{\theta}_\mathrm{optimal} = \{0.94, 200.74, 0.79\}^{\mathrm{T}}$. 
For the current selection of free parameters, all the derivatives have analytic solutions as the derivative of the covariance of the LSS component \citep{reischke_cosmological_2023} scales with $2\mathcal{A}$, with $\mathcal{A}$ being the prefactor in \Cref{eq:DM_LSS_v2}. Similarly, the derivatives of the log-normal host covariance are trivial. If we are to fit a more complex model in the future, one would have to calculate those derivatives numerically or, for more stable results, use a differentiable code. This could, for example, be achieved by emulating the important quantities.

\section{Inference}
\label{sec:inference}
\begin{table}
\renewcommand{\arraystretch}{1.2}
    \centering{
    \begin{tabular}{cccc}
      Parameter   & $\mathcal{A}$ & $\mathrm{DM}_\mathrm{host}$ & $\sigma_\mathrm{LN}$  \\
      \hline\hline
      Prior range   & $[0,3]$ & $[0,1500]$ &$(0, 1.5]$  \\
      Fiducial & $1$ & $200$ & $0.35$
    \end{tabular}}
    \caption{Parameters fitted in the inference with prior ranges (flat) and fiducial values for the data compression.}
    \label{tab:parameters}
\end{table}
Given the simulation pipeline described in the previous section, we are now in the position to infer the posterior of the parameters summarised in \Cref{tab:parameters}.
Inference refers to determining the parameters that best describe the observation. Typically, we have a model based on physical laws, observations from surveys and a likelihood. The latter is then combined with the prior to yield the posterior. The most popular approach to such an inference process has been the Markov Chain Monte Carlo (MCMC) within the Bayesian framework. MCMC methods create samples from the posterior. However, modelling a likelihood that captures the full complexity of the physical processes is one of the most complex parts of any traditional analysis, with cases arising where an analytical likelihood is not accessible. 

\subsection{Simulation-based inference}

Following the development of NNs, a suite of algorithms has been proposed recently that estimate the posterior without access to a likelihood. All of these methods come under the umbrella of SBI or likelihood-free inference (LFI), which introduces a parameterised density estimator that learns from the joint distribution of the data-parameter pair \citep{cranmer2020frontier}. In this section, we briefly introduce the density estimation mentioned in \citep{papamakarios2016fast} called sequential neural posterior estimation (SNPE-A) and summarise its subsequent developments in SNPE-B \citep{lueckmann2017flexible}, SNPE-C or APT \citep{greenberg2019automatic} and an improvement on APT called TSNPE \citep{deistler2022truncated}. Finally, we will be using the TSNPE algorithm in our analysis.

The fundamental idea behind all of these methods is to approximate the posterior, $p(\boldsymbol{\theta}|\boldsymbol{\mathrm{d}})$, with a conditional density estimator $q_{\phi}(\boldsymbol{\theta}|\boldsymbol{\mathrm{d}})$, where $\phi$ are the trainable parameters. Typically, the parameters, $\boldsymbol{\theta}$, are sampled from the full prior range, which is inefficient. \review{There exist two schools of SBI. Amortised methods yield a posterior that can be utilised for various observations without the need for retraining, whereas sequential methods concentrate the inference on a specific observation to enhance simulation efficiency. Note that these two methods are not strictly mutually exclusive. Even though sequential SBI is tailored to a specific observation, the resulting posterior can be utilised in an amortised manner. If the new observation for the amortised case is not significantly different from the one used in sequential training, the older network can be applied with additional training.}

Here, we focus on the sequential-(S)NPE methods, which refer to using a proposal as a subset of the prior, $\Tilde{p}\boldsymbol{\theta}) \subseteq p(\boldsymbol{\theta})$. Initially, the proposal is equal to the prior and in subsequent rounds, we update the proposal with the approximate posterior, improving the efficiency of the inference process. The predicament is that this procedure does not recover the true posterior, but rather an approximate posterior:
\begin{align}
    \Tilde{p}(\boldsymbol{\theta}|\boldsymbol{\mathrm{d}}) = p(\boldsymbol{\theta}|\boldsymbol{\mathrm{d}}) \ \frac{\Tilde{p}(\boldsymbol{\theta})p(\boldsymbol{\mathrm{d}})}{p(\boldsymbol{\theta})\Tilde{p}(\boldsymbol{\mathrm{d}})}\;,
\end{align}
where $p(\boldsymbol{\theta}|\boldsymbol{\mathrm{d}})$ is the true posterior, $\Tilde{p}(\boldsymbol{\theta}|\boldsymbol{\mathrm{d}})$ is the approximate posterior, $\Tilde{p}(\boldsymbol{\theta})$ is the proposal, $p(\boldsymbol{\theta})$ is the prior and $\Tilde{p}(\boldsymbol{\mathrm{d}})$ is the evidence. The approximation recovers the true posterior only when the proposal is equal to the prior. 

SNPE methods address this issue with a parameterised density estimator, $q_\phi(\boldsymbol{\theta}|\boldsymbol{\mathrm{d}})$, that learns from the joint distribution of the data and parameter
\begin{align}
\label{eq:unnormalised_posterior}
    \Tilde{p}(\boldsymbol{\theta}|\boldsymbol{\mathrm{d}}) \propto  q_{\phi}(\boldsymbol{\theta}|\boldsymbol{\mathrm{d}}) \ \frac{\Tilde{p}(\boldsymbol{\theta})}{p(\boldsymbol{\theta})}\;.
\end{align}
The problem to solve is the extra ratio, $\Tilde{p}(\boldsymbol{\theta})/p(\boldsymbol{\theta})$. The way each algorithm addresses this issue is what differentiates the various SNPE methods. For instance, SNPE-A maintains a closed-form solution by restricting the choice of the prior, proposal and density estimator to either a uniform or Gaussian distribution. The training is carried out by minimising the loss function or the negative log-likelihood of the joint probability of the data and parameter defined via
\begin{align}
    \mathcal{L}(\phi) \coloneqq \mathrm{min} \ \mathbb{E}_{\boldsymbol{\theta} \sim \Tilde{p}(\boldsymbol{\theta})} \ \mathbb{E}_{\boldsymbol{\mathrm{d}} \sim p(\boldsymbol{\mathrm{d}}|\boldsymbol{\theta})} \left[- \ \mathrm{log} \ q_{\phi}(\boldsymbol{\theta}|\boldsymbol{\mathrm{d}})\right]\;.
\end{align}
Other methods like SNPE-B offer a solution by including the ratio inside the loss function, whereas SNPE-C incorporates the ratio into the density estimator and finally, APT normalises the density estimator by using uniform subsets of the prior called atoms. This development allows any arbitrary distribution choice for the prior, proposal and density estimator. For an extensive technical review of the shortcomings and the subsequent developments of the various SNPE methods, the reader is referred to the following articles and the references therein \citep{durkan20a, lueckmann21a, xiong_2023_SNPE_B_problems}.

The latest development, TSNPE, builds upon the APT method and truncates the prior based on the $1-\epsilon$ mass ($\epsilon$ being an arbitrary margin) of the highest probability region (HPR) of the approximate posterior. In other words, the subsequent rounds only consider the prior range, which is most probable to produce the posterior. This prescription maintains the flexibility of the distribution choice while also accounting for the posterior mass outside the prior boundaries, an issue of the APT method \citep{deistler2022truncated}.

As the HPR of the approximate posterior contains the information on the joint probability of the data and parameter in each round, the truncation is data-driven. TSNPE also results in faster overall convergence.

The various SNPE methods tackle posterior estimation by either changing the loss function or restricting the choice of distribution. Without a general solution, the choice of algorithm affects the performance and accuracy of the analysis. Our tests reveal that TSNPE, due to the truncation, is faster and allows for a more flexible distribution choice for prior and density estimators as compared to SNPE-A. 
As the field of NNs is projected to experience rapid growth, the choice of the best algorithm is expected to change as well; refer to \citep{cranmer2020frontier} for a detailed review of the status of SBI at the beginning of this decade. 

\subsection{Implementation}
\label{subsec:implementation}

The simulated data in our case are the compressed DM values of FRBs as described in Section \ref{subsec:compression}. \review{We use uniform priors and the ranges are defined in \Cref{tab:parameters}, where the limits are motivated by the results from \citet{hagstotz_2022_new, reischke_consistent_2022}. The priors are all wide enough, especially for $\mathcal{A}$ and $\mathrm{DM}_\mathrm{host}$, that the data have to constrain them. Since $\sigma_\mathrm{LN}$ controls the shape of the distribution, it will be more dominated by the prior as this would require more FRBs. However, increasing $\sigma_\mathrm{LN}$ above the provided prior bound would create a host PDF whose peak is at very low DM; this has not been observed in simulations \citep{theis_galaxy_2024,medlock_probing_2024}.} For the analysis, we use 12 FRBs with host identification as given in \Cref{tab:frb_cat}, where, amongst others, we list the observed DMs and their references. 

The forward simulations for the fiducial case of log-normal density field and log-normal host contribution are carried out at a resolution of $N_\mathrm{side} = 4096$. The choice is so that DM contributions from small scales ($\ell \sim 10^4$) are accounted for. This is essential for an unbiased analysis, as we shall discuss. A good consistency check here is the comparison of the numerical covariance we get from the simulations to its analytical counterpart \citep{reischke_cosmological_2023}. The diagonal elements of the analytic covariance matrix show a monotonic behaviour between the DM and the redshift of FRBs. For the numerical covariance, we choose $N_\mathrm{side}=4096$ to match this monotonous nature and magnitude of the diagonal elements, i.e. the DM field variance.

In each round of TSNPE, the density estimator learns the mapping between the simulated data and the corresponding parameters. The observed DM is introduced post-training, which helps evaluate an approximate posterior for that round. This approximate posterior is then utilised to calculate the HPR region and the prior range for the next round is truncated based on that. As the number of rounds is increased, the approximate posterior converges to the true posterior. Introducing the observed DM is necessary to propel the truncation in the direction of the true posterior. {With parameters, simulated data and observation, these are the three components required for the SBI analysis. We implement this using the \texttt{LtU-ILI} package for a quick setup with different neural embedding \citep{ho2024ltu}.} We have used the TSNPE algorithm for a total of 1500 simulations in 10 rounds. The initial simulations contain the full prior range and the subsequent truncation is only influenced by the data. As \texttt{LtU-ILI} does not natively support TSNPE yet, we run TSNPE, save the data and then use it. The choice of the number of simulations in each round is dictated by the number of epochs the NN takes for training. We limit the number of epochs to 50 to avoid overfitting and 150 simulations in each round provide ample data for the density estimator $q_{\phi}(\boldsymbol{\theta}|\boldsymbol{\mathrm{d}})$ to approximate the posterior by adjusting the trainable parameter $\phi$.

As for the NN architecture, we use two different types: mixture density network (MDN) and masked autoregressive flows (MAF). These architectures aim to produce an arbitrary yet tractable PDF with trainable parameters to be used as the density estimator. The complexity of data mandates arbitrariness, while the backpropagation in the training process requires tractability. MDN, as the name suggests, is the weighted sum of many Gaussian distributions defined as \citep{bishop1994mixture}
\begin{align}
    q^\mathrm{MDN}_\phi(\boldsymbol{\theta}|\boldsymbol{\mathrm{d}}) \coloneqq \sum_k \alpha_k \mathcal{N}(\boldsymbol{\theta}|\mathrm{m}_k, \boldsymbol{\mathrm{S}}_k) \;,
\end{align}
where $\mathrm{m}_k$ and $\boldsymbol{\mathrm{S}}_k$ are the mean and covariance of the $k$-th Gaussian. This type of density estimator is malleable and can be transformed into any arbitrary distribution by adjusting the weights, means and standard deviations. While flexibility is a property we seek, a value of $k$ too large can lead to overfitting. Masked Autoregressive Flows \citep[MAFs,][]{papamakarios2017masked}, on the other hand, combine two different architectures called Normalizing Flow \citep[NF,][]{rezende2015variational} and Masked Autoencoder for Distribution Estimation \citep[MADE,][]{germain2015made} into one. In NF, a normal distribution is subjected to $k$ invertible transformations, called flow, resulting in an arbitrary complex distribution. These transformations are based on the conditional probability scheme described in MADE. Combining NF and MADE, MAF can produce arbitrary distribution optimised for density estimation. For technical details, the reader is referred to the respective articles. The presence of MAF along with MDN (7:3 weight ratio) ensures the model has a better scaling with dimensionality and learns any multimodal, non-Gaussian features of the data while avoiding overfitting.

\review{For the MDN embedding, we use 50 hidden layers with 10 components in each layer, while the MAF net is constructed using 20 hidden layers with 8 transformations in each layer. The network is trained to minimise the negative log-likelihood or the Kullback–Leibler (KL) divergence between the approximate and the true posterior. For our case, this minimisation is achieved with $\sim$ 800 simulations. When the training is complete, we acquire a mapping (effectively a KDE) from the joint space of parameters to the distribution $p(\boldsymbol{\theta}|\boldsymbol{\mathrm{d}})$ without the usage of an explicit likelihood function. This posterior can then be sampled by standard MCMC techniques \citep[here we use \texttt{emcee} introduced in][]{foreman-mackey_emcee_2013}. We use 3 walkers and they all converge consistently.}

\section{Validation}
\label{sec:validation}
After the completion of the SBI pipeline, we carry out two validation tests, one for the model under consideration and one for its posterior. For the posterior, we use the multivariate coverage test called Tests of Accuracy with Random Points \citep[TARP,][]{pmlr-v202-lemos23a} and for the model, the accuracy is measured via the goodness of fit or $\chi^2$-test. While TARP validates the SBI, the $\chi^2$-test is used to investigate if there are preferred models. In the following, we describe the two validation techniques.

\subsection{Coverage Test}
To check the consistency of the posterior distributions, we use the multivariate coverage test called TARP. Coverage probability measures how frequently the estimated posterior contains the true parameter value and can be used to check the consistency of the posterior \citep{guo2017calibration, hermans2021trust}. Once we have obtained the posterior, TARP measures the expected coverage probability of random posterior samples within a given credibility level of the learned posterior empirically.
In more technical terms, for any instance of the prior $\boldsymbol{\theta}^\star$ and the corresponding simulated data $\boldsymbol{\mathrm{d}}^\star$, samples are drawn from the posterior as $p(\boldsymbol{\theta}|\boldsymbol{\mathrm{d}^\star})$. Then, a circle is drawn centred at a random reference point $\boldsymbol{\theta}_\mathrm{r}$ with radius $|\boldsymbol{\theta}^\star - \boldsymbol{\theta}_\mathrm{r}|$ and the fraction of posterior samples inside the circle is calculated. A higher fraction implies the posterior is more accurate. The fraction is calculated multiple times to get a statistical estimate of the coverage. 

\begin{figure}
    \centering
    \includegraphics[width=\linewidth]{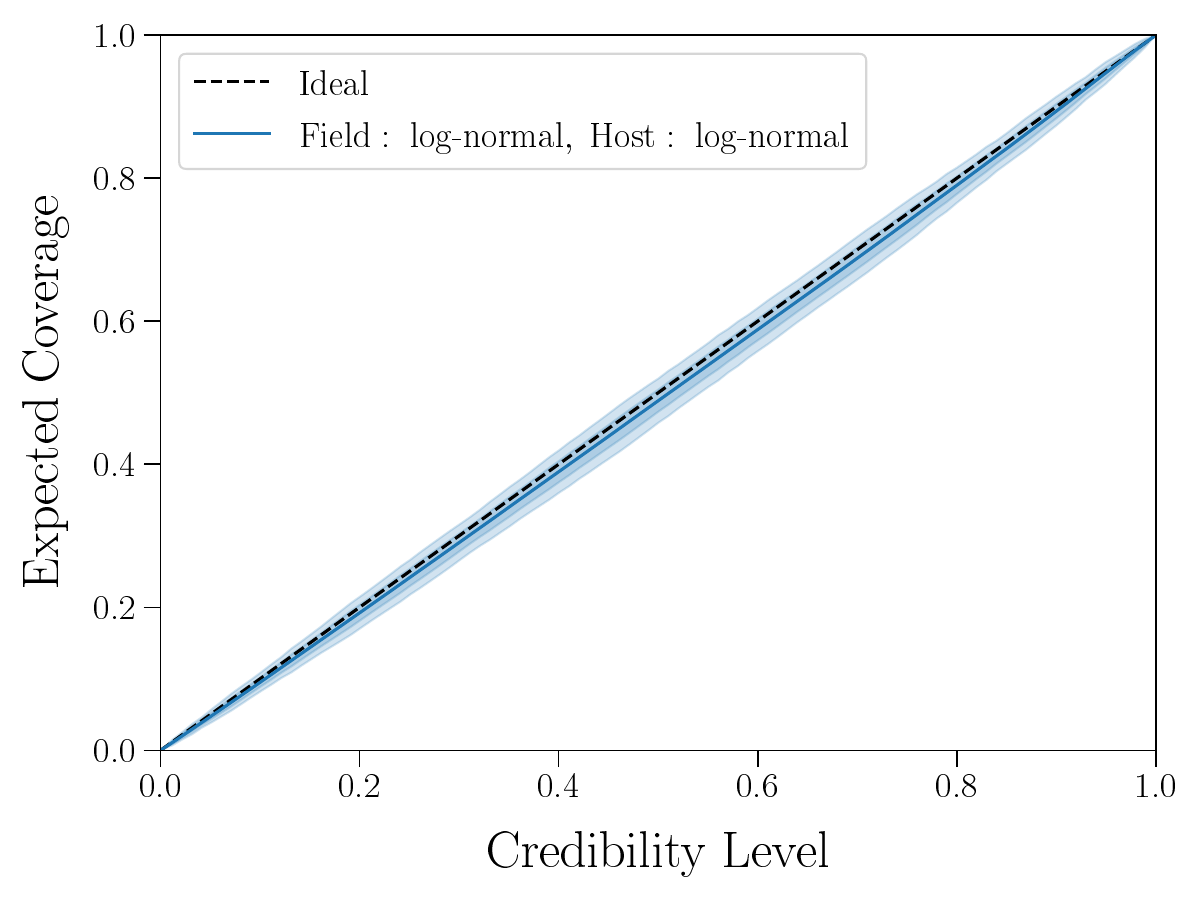}
    \caption{TARP coverage for the fiducial case of log-normal field and log-normal host is shown with $N_\mathrm{side} = 4096$. Credibility level refers to the fraction of the total PDF of the final posterior in \Cref{fig:contour_4096} being considered, which intuitively goes from 0 to 1. The expected coverage is the fraction of the posterior samples with a lower posterior probability than the best estimate for that corresponding credibility. The diagonal line is the ideal relation, as increasing the credibility level linearly increases the expected coverage. For example, at a 50\% credibility level, we expect at least 50\% of the samples to have coverage; anything below 50\% implies the samples are not well covered, i.e. the posterior is biased and similarly, a value greater than 50\% implies that the posterior is conservative. The dark and light-shaded regions are the one and two-sigma error bars from 100 bootstrappings. Our model, the solid blue line, is accurate as the ideal line is within the $1\sigma$ error bar.}
    \label{fig:tarp_4096}
        \vspace{.3cm}
\end{figure}

This random sampling is necessary to validate the static HPR region in TSNPE, which is prone to having blind spots. 
We use the learnt posterior to draw 1000 samples at random parameter points for the TARP test implemented within the \texttt{LtU-ILI}. \review{All three parameters in our analysis, $\mathcal{A},\, \mathrm{DM}_\mathrm{host}$ and $\sigma_\mathrm{LN}$, are kept free and their prior ranges are defined in \Cref{tab:parameters}.} We also bootstrap the test 100 times at each credibility level, considering the process is random by definition. \review{We elaborate on the result of the coverage test as shown in \Cref{fig:tarp_4096} in Section \ref{sec:results}.}

\subsection{Goodness-of-Fit}
The typical goodness of fit test can not be used in the SBI framework, as it requires an analytical likelihood. Here, we use the implementation of the $\chi^2$-test \citep{gelman1996posterior} described in \citet{von_wietersheim-kramsta_kids-sbi_2024}.

After finding the best-fit values for the parameters of interest, $\boldsymbol{\Theta}_*$, we sample $n$ noise realisations at this set of parameters and define the $\chi^2$ as
\begin{equation}
\chi^2(\boldsymbol{t}_i|\boldsymbol{\Theta}) \coloneqq(\boldsymbol{t}_i - \mathrm{E}[\boldsymbol{t}_*|\boldsymbol{\Theta}_*])^\mathrm{T}{(\mathrm{Cov}(\boldsymbol{t}_*|\boldsymbol{\Theta}_*)})^{-1}(\boldsymbol{t}_i - \mathrm{E}[\boldsymbol{t}_*|\boldsymbol{\Theta}_*])\;,
\end{equation} 
where E denotes the expectation value and Cov is the covariance at the best-fit cosmology. $\boldsymbol{t}_i$ is the score-compressed summary statistic of the $i$-th realisation. 
Specifically, we create 1000 new forward simulations at the best-fit values to calculate the mean data vector and the covariance. The $\chi^2$ values from the simulations are plotted as a histogram and compared with the $\chi^2$ value for the observed DM. \review{The resulting figure is shown in \Cref{fig:chi2_4096} and explained in Section \ref{sec:results}.} The accuracy of the model is assessed through the probability mass of the simulated $\chi^2$ that lies beyond this observed $\chi^2$, which we call the probability to exceed (PTE).

\begin{figure}
    \centering
    \includegraphics[width=\linewidth]{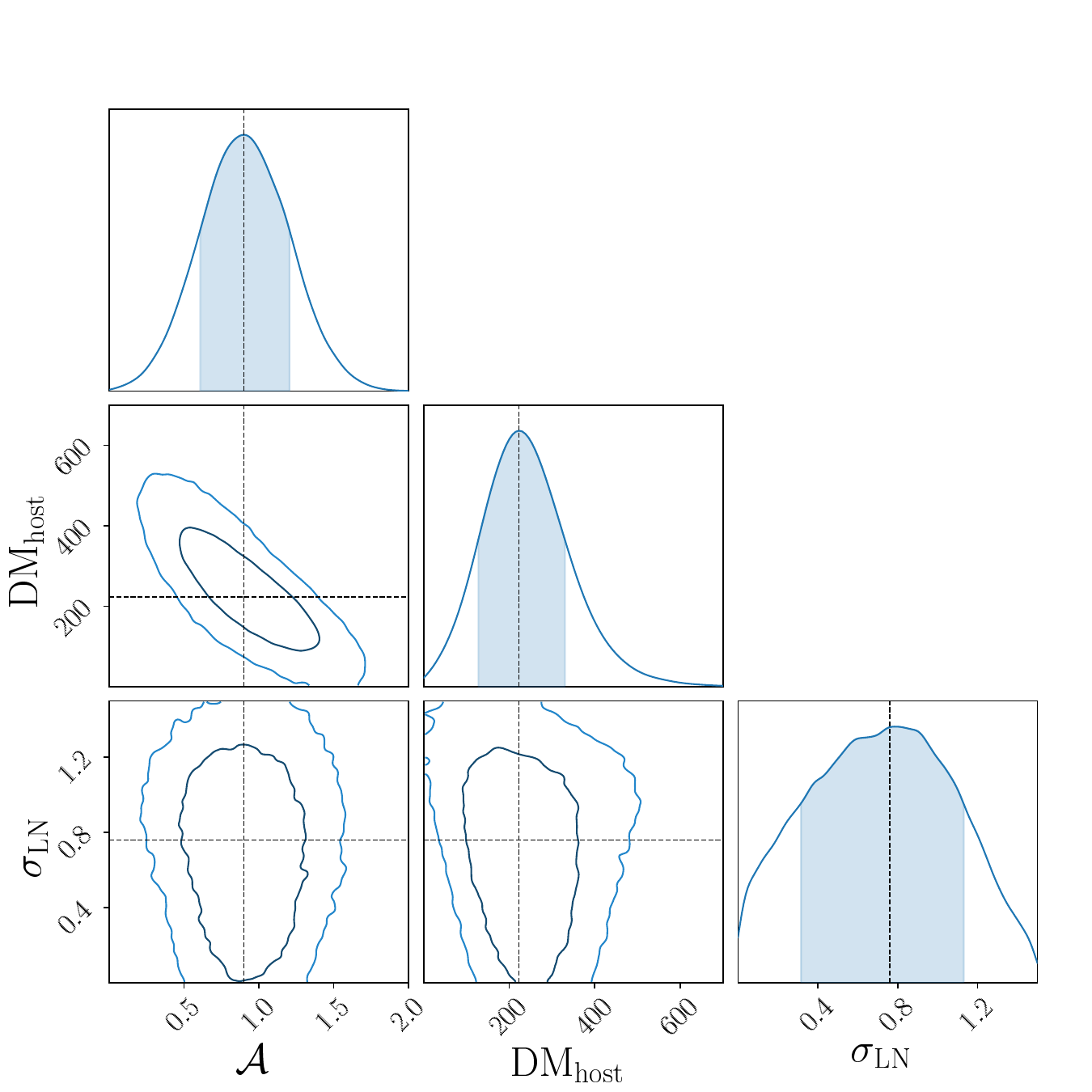}
    \caption{Marginalised contour plot for the fiducial model in our analysis, i.e. log-normal field and log-normal host. The shaded areas represent the $1\sigma$ error margin around the dotted line or the mean. The best-fit values given as $\mathcal{A} = 0.89^{+0.31}_{-0.29}, \mathrm{DM}_\mathrm{host}= 223^{+108}_{-95}\;\mathrm{pc \;cm}^{-3}\;\mathrm{and}\;\sigma_\mathrm{LN}= 0.76^{+0.37}_{-0.44}$.}
    \label{fig:contour_4096}
    \vspace{.3cm}
\end{figure}

\section{Results}
\label{sec:results}
In this section, we present the major results of our analysis. We consider three parameters for the inference. The first one is the amplitude of the DM-$z$ relation in \Cref{eq:DM_LSS_v2} as a scale. This encapsulates the $\Lambda$CDM parameters, for which we use the Planck cosmological results from \citep{planck_collaboration_planck_2020}, i.e. $\mathcal{A}=1$ refers to the input cosmology and this $\mathcal{A}$ is varied in different simulations. The motivation behind only varying the amplitude of the DM-redshift relation is that the number of FRBs currently available is not able to discriminate between different expansion histories. In other words, we can only aim to infer $H_0$ but not $E(z)$ \citep[see e.g.][]{macquart2020census,hagstotz_2022_new,james_measurement_2022}.
The remaining two parameters are the median and the scale of the log-normal host distribution, which we denote as $\mathrm{DM}_\mathrm{host}$ and $\sigma_\mathrm{LN}$.  

As for the model, there are two components in our simulations that vary. First, the $\mathrm{DM}_\mathrm{LSS}$ part is modelled using either a log-normal or a Gaussian realisation of the underlying electron density field via \texttt{GLASS}. These are denoted by `Field: log-normal' and `Field: Gaussian' respectively. Similarly, we have three host distributions: log-normal, truncated Gaussian and Gamma. With this setup, we have six possible combinations of field and host. That being said, we use a log-normal field and a log-normal host distribution as our fiducial setting. For this, we will discuss and validate the results in detail and use the high resolution setting of $N_\mathrm{side} = 4096$. For the other combinations, we use a lower resolution $N_\mathrm{side} = 512$ in order to limit computation time. In the following, we start with the fiducial model and then observe how the changes in field and host affect the results.

\begin{figure}
    \centering
    \includegraphics[width=\linewidth]{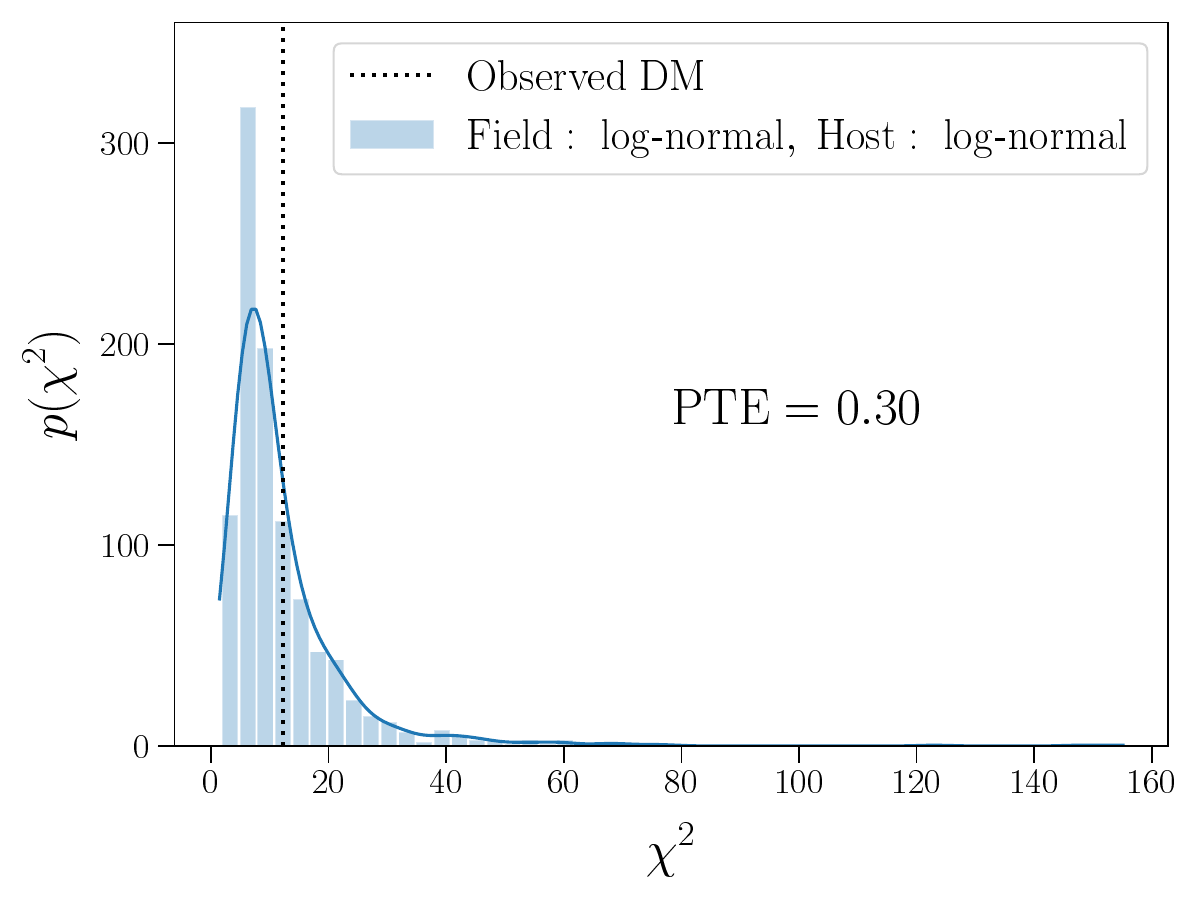}
    \caption{Goodness-of-fit test that shows the distribution of $\chi^2$ values. The solid blue line is obtained by a smoothing kernel applied on the histogram built from 1000 forward simulated DM values at the best-fit value of \Cref{fig:contour_4096} with $N_\mathrm{side} = 4096$. The dotted vertical black line is the $\chi^2$ value for the observed DM. The probability mass beyond this line, PTE, is 0.3.}
    \label{fig:chi2_4096}
    \vspace{.3cm}
\end{figure}

\subsection{Fiducial model and validation}
For the fiducial model, we assume the electron density field in \texttt{GLASS} to be log-normal and sample the host DM from a log-normal distribution. The decision is primarily motivated by results from simulations \citep[see e.g.][]{zhang_intergalactic_2021, theis_galaxy_2024}. In the subsequent sections, we present the comparisons of the six different models which support our decision.

With the simulator set to the desired configuration and priors defined in \Cref{tab:parameters}, the SBI pipeline is run with the TSNPE algorithm, which returns the posterior distribution for the free parameters. In \Cref{fig:contour_4096}, we show the marginalised contour plots using \texttt{ChainConsumer} \citep{Hinton2016}. The forward simulations have the resolution parameter $N_\mathrm{side}=4096$. As mentioned at the beginning of this section, there are three parameters that we constrain; $\mathcal{A}$ as a scale for the input cosmology, $\mathrm{DM}_\mathrm{host}$ as the median and $\sigma_\mathrm{LN}$ as the scale of the log-normal host distribution. The value for the scale parameter $\mathcal{A}$ is $0.89^{+0.31}_{-0.29}$ at $68\%$ confidence, which is consistent with unity or the input fiducial Planck cosmology. The median and the scale of the log-normal host along with their $68\%$ confidence intervals are given as $\mathrm{DM}_\mathrm{host} = 223^{+108}_{-95}\;\mathrm{pc \;cm}^{-3}$ and $\sigma_\mathrm{LN}= 0.76^{+0.37}_{-0.44}$. \review{The scale parameter essentially controls the skewness of the distribution. However, for a more intuitive comparison of the fiducial host (log-normal) to the non-fiducial host (Gaussian and Gamma) distributions, we convert $\sigma_\mathrm{LN}$ into the standard deviation from the definition of the variance in \Cref{eq:host:lognormal_pdf}.} Then, we write $\sigma = 263_{-113}^{+91}$. {As expected, we find a strong anti-correlation between $\mathcal{A}$ and $\mathrm{DM}_\mathrm{host}$ since they both enter the observed DM in an additive fashion. The width of the host contribution is not degenerate with the other parameters, as it rather reduces the error on each measurement than changing the signal.}

Our constraints of $\sigma_\mathrm{LN}$ seem to be slightly prior driven, as can be seen from the posterior hitting the prior boundaries of $\sigma_\mathrm{LB}$ in \Cref{fig:contour_4096}. However, there is clearly some constraining power in the data on the shape of the host distribution. In general, we find excellent agreement with previous work \citep{macquart_census_2020,james_measurement_2022,hagstotz_2022_new,khrykin2024flimflam}. \review{The most direct comparison of our work is \citet{reischke_consistent_2023-1}, where the scale $\sigma$ is kept constant. In our case, the mean of $\mathcal{A}$ improves by $\sim 31\%$ and the errors on $\mathrm{DM}_\mathrm{host}$ shrink by $\sim 22 \%$ while the mean agrees at $\sim 62\%$.}  

Now that the posterior distribution is evaluated, we first assess its accuracy via the TARP coverage in \Cref{fig:tarp_4096} with the solid blue line. The shaded blue regions are the 1 and $2\sigma$ error bars from the 100 bootstraps. The region above the diagonal dotted line is called under-confident or conservative, while the region below suggests overconfidence or bias. As the diagonal dotted line is within the dark-shaded region throughout the whole range of credibility, this means that the learnt posterior is an accurate representation of the true posterior and is not biased.

Additionally, we check for the goodness of fit of the model, and the result is shown in \Cref{fig:chi2_4096}. The histograms are the $\chi^2$ values from our simulations at best-fit values of the parameters, and the vertical line corresponds to the $\chi^2$ evaluated at the observed DM. The PTE for the fiducial model is 0.3 (corresponding to a $p$-value), \review{which means that the observation agrees with the majority of the probability mass and is not an outlier, implying that the model is a good fit.} We can thus concur that the `log-normal field and log-normal host' model is an excellent fit for the DM of FRBs.

\begin{figure}
    \centering
    \includegraphics[width=\linewidth]{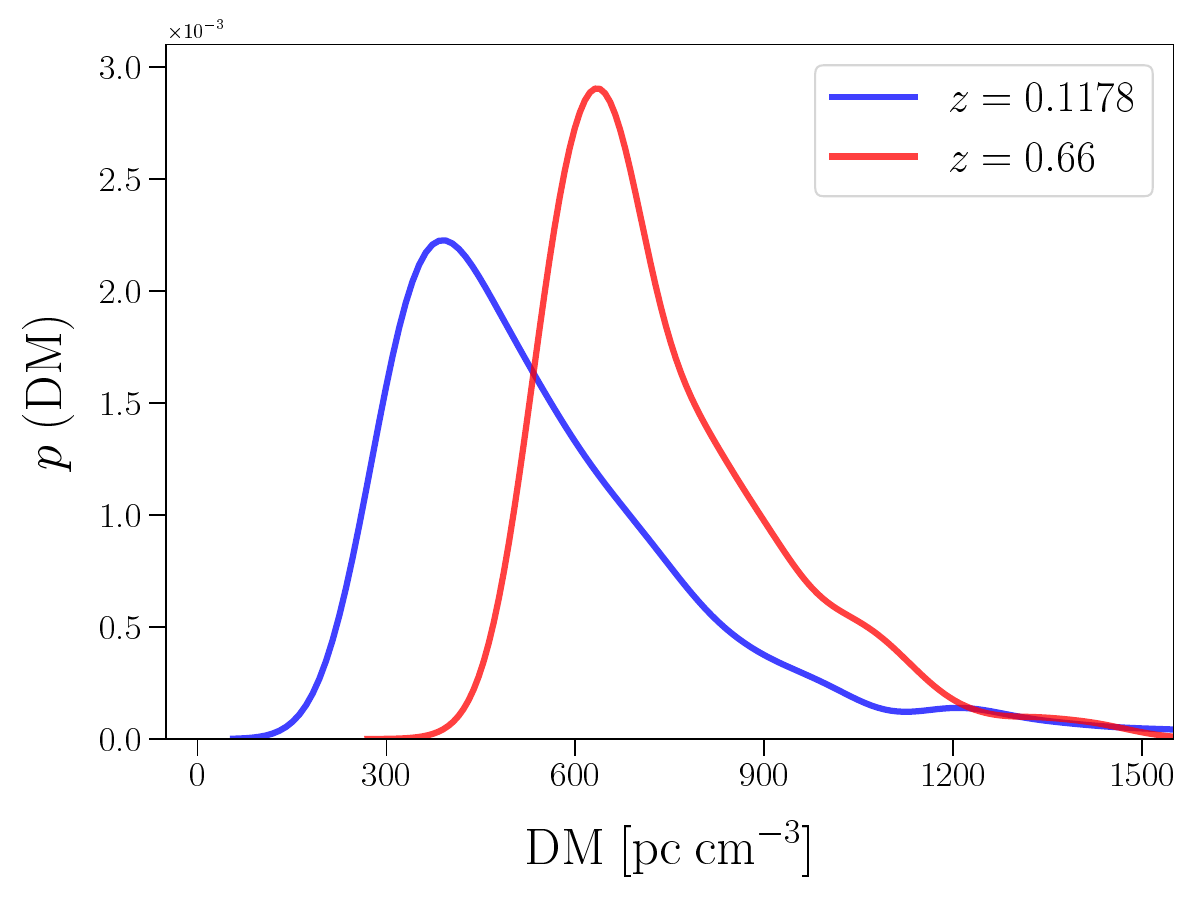}
    \caption{Likelihood of the DM for two FRBs at different redshifts as evaluated from the simulations. The blue line is for a lower redshift of $z=0.1178$, while the red line is for $z=0.66$. The non-Gaussian nature is obvious, along with the distinct long tail, which accounts for the high DM values originating from the disk of the host halo.}
    \label{fig:likelihood}
    \vspace{.3cm}
\end{figure}

With the SBI framework, all the correlations are taken into account without defining any likelihood function. Nonetheless, we can still recover the likelihood from the simulated DMs for the best-fit values. We, therefore, run our pipeline 1000 times at the best-fit value and plot the corresponding distribution of DM.
In \Cref{fig:likelihood}, we show the DM likelihoods for two FRBs with varying redshift, $z = 0.1178$ and $z=0.66$. The first observation is the non-Gaussian nature of the distribution, with a tendency towards higher DMs. This behaviour emerges from the host and LSS DM contribution. 
If the electron density of the host halo where the FRB progenitor resides is high, the signal experiences higher dispersion. That seems to be especially the case for FRB 20190520B, with an observed $\mathrm{DM} = 1202 \;\mathrm{pc \;cm}^{-3}$ at a relatively lower redshift of $z=0.241$ as shown in \Cref{tab:frb_cat}. The high DM is due to the local contribution of the dwarf host galaxy, identified as J160204.31-111718.5 \citep{ocker2022largeDM, niu_persistent_2022, yan2024simultaneous}. The implication is that a Gaussian likelihood assumption in the inference can introduce biases, as it does not consider the physical effect of variable travel distances and local environments of the host. Furthermore, we observe that the mean shifts towards higher DMs in keeping with the DM-$z$ relation. We also see that the width of the distribution seems to be getting slightly smaller. This might be counter-intuitive at first, as the LSS component should increase with redshift, hence causing a broader distribution. However, the host contribution is scaled down with $1/(1+z)$ in physical coordinates. \review{At the same time, the density of the electron density field reduces as well with increasing redshift, the DM itself, is still an additive effect.} Thus, those two counter-acting effects can lead to a decreasing width of the likelihood as a function of redshift. The strength of this effect itself depends on the parameters of the model, in particular on the parameters of the host contribution. To understand this a bit better, let us consider a simple toy model. We call the variance of the host at redshift zero $\sigma^2_\mathrm{host,0}$ and label the variance of the LSS at $z=1$ as $\sigma^2_\mathrm{LSS,1}$. For linear structure growth in a matter-dominated Universe, one has $\delta_\mathrm{e}\propto a$. The LSS variance scales roughly linear with redshift in this case \citep{reischke_cosmological_2023}. Therefore, the total variance is given by
\begin{equation}
\label{eq:scaling_z}
    \sigma^2(z) = \frac{\sigma^2_\mathrm{host,0}}{(1+z)^2} + \sigma^2_\mathrm{LSS,1}z\,,
\end{equation}
which has a minimum at $z = (2\sigma^2_\mathrm{host,0}/\sigma^2_\mathrm{LSS,1})^{1/3}$. Plugging in the values we find in our analysis, the minimum arises around $z=1$.
 This simple model shows that the variance of the likelihood can initially decrease and then rise again at larger redshifts. This is exactly what we observe in \Cref{fig:likelihood}. If FRBs at larger redshifts, $z\gtrsim 1$, become available, an increase in the width of the likelihood should become visible.

\begin{figure}
    \centering
    \includegraphics[width=\linewidth]{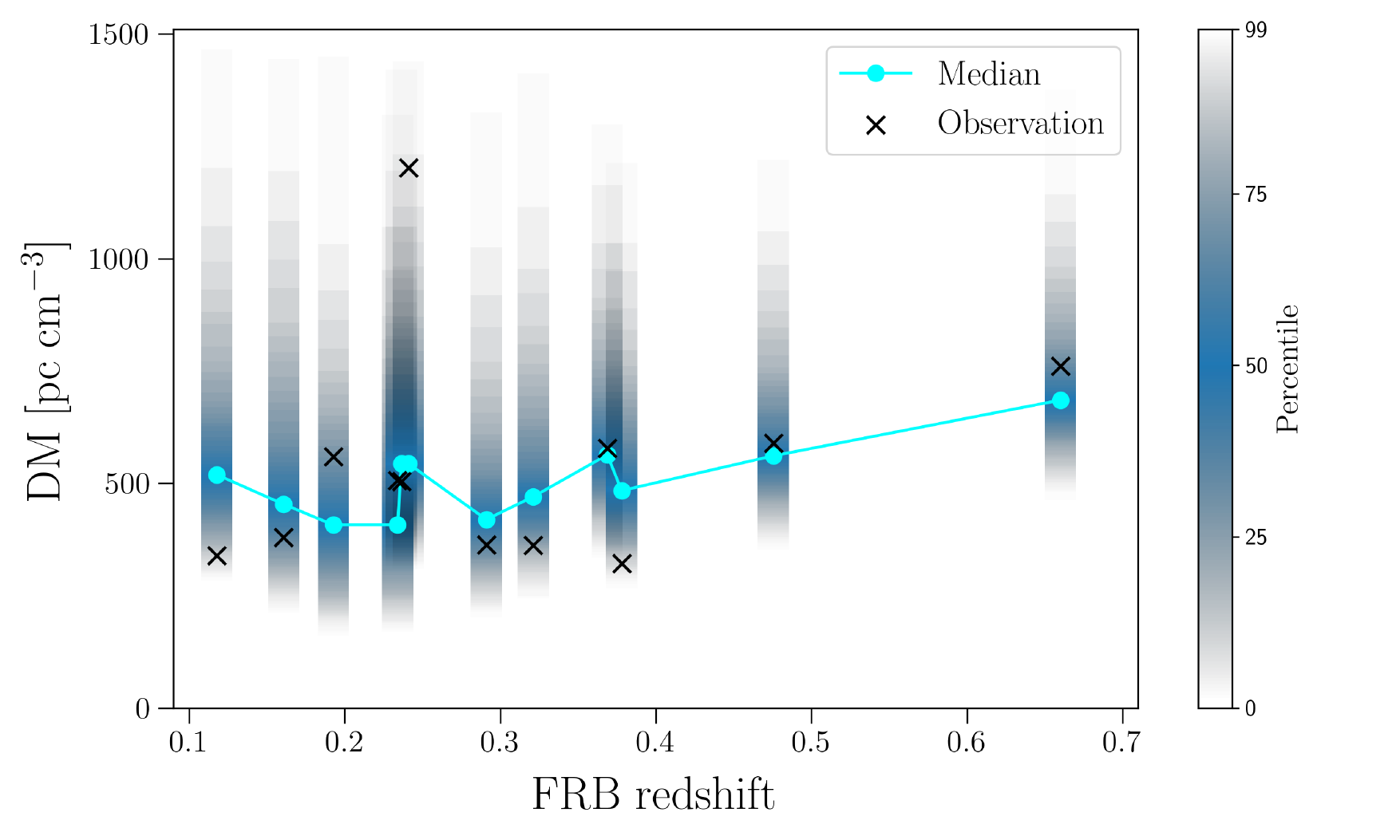}
    \caption{Model prediction with likelihoods for the best-fit model from \Cref{fig:contour_4096}. The median (cyan dots) and the observed DM values (black cross) are overlaid on the DM values from the 1000 forward simulations of our fiducial model with log-normal field and log-normal host distribution. The colour bar is representative of the percentiles of the DM values for each FRB, where the darkest shade is at the median and falls off on both sides.}
    \label{fig:all_data}
        \vspace{.3cm}
\end{figure}

Finally, we show the model prediction at the global maximum posterior together with the likelihood for all data points in \Cref{fig:all_data}. Note that the model prediction includes all components of the DM in \Cref{eq:dispersion_measure_contributions}. Therefore, the relation between DM and redshift is not necessarily monotonous. 
The medians of the simulated DMs are shown in cyan dots for the redshifts of the FRBs. The colour bars represent the different percentiles. They include the observed data (black cross) within the darker shades, i.e. 25 to 75 percentiles. We would like to remark that those percentiles are, in principle, correlated due to the LSS contribution. However, this correlation is not important for the number of FRBs considered here \citep{reischke_cosmological_2023}. It is again apparent that the long tails of the likelihood are required to explain the large scatter in DM values. We furthermore investigated the response of the inference to leaving out FRB20190520B, which has a DM > 1000 but is located at a low redshift. Our findings show that the host contribution responds with a lower median and width by roughly 10 and 20 percent, respectively. This, of course, is still fully consistent within the error bars. It shows, however, that this particular FRB increases the values of the inferred host contribution.
Lastly, we recover the mean DM-$z$ relation as well by connecting the cyan dots.

\review{\subsection{Non-fiducial model and comparison}
In the following, we move from the fiducial model and compare all the possible combinations of LSS field and host distributions. With our established pipeline, this requires changing only the sampling function from `log-normal' to `Gaussian', for example. We also reduced the resolution parameter, $N_\mathrm{side}$, to 512 to decrease the computation time. This, however, produces bias or overconfidence in the TARP coverage test. We observe that this is from a combined effect of $(i)$ not including the DM contribution from smaller scales and $(ii)$ smoothening of pixels from interpolation, both of which result in lower variance of the LSS component. We elaborate more on this effect in appendix \ref{app:nsideeffect}. We note that the bias at low $N_\mathrm{side}$ is present even for the fiducial case and is minimised by increasing the resolution. This allows us to compare the different combinations directly and any comparative conclusion we draw is agnostic to the choice of $N_\mathrm{side}$.}

\subsubsection{Variations in the LSS component}
Our fiducial model consisted of a log-normal realisation of the LSS component along with a log-normal host model. Next, we change the LSS component to a Gaussian distribution and observe how the results are affected, which can be done trivially with our pipeline. To that end, the SBI pipeline, as described in Section \ref{subsec:implementation}, is applied by changing the field in \texttt{GLASS}. For the comparison, we reduce the resolution of the simulations to $N_\mathrm{side}=512$. As we have established that the high-resolution run faithfully reproduces the real posterior distribution and is a good fit to the data in the last section, reducing the resolution to $N_\mathrm{side} = 512$ makes a quantitative comparison while also requiring less computational resources. We also rerun the fiducial case at the same resolution as well to make a quantitative comparison. It should be noted that the lower resolution effectively reduces the field variance of the LSS component. In appendix \ref{app:nsideeffect}, we discuss the effect of reducing $N_\mathrm{side}$ on the coverage tests. It can be seen that the resulting posterior estimates are slightly biased when $N_\mathrm{side}$ is lowered, reducing the variance of the LSS component. Since we have established unbiased estimators at high resolution and checked that the constraints and $\chi^2$-test are not affected, this does not hamper our analysis when comparing different models.

The contours are shown in \Cref{fig:contour_plots} and the numerical values of the means and the $1\sigma$ errors of the parameters are presented in \Cref{tab:best_fit_all}. As can be observed, there is a significant overlap among the values. Only by observing the contour plots we cannot distinguish the models. Hence, we rely on the $\chi^2$-tests, shown in \Cref{fig:chi2}, respectively. Even then, judging from these figures, there is no discernible difference between the Gaussian and log-normal LSS components. All the variations seem to be good fits according to the likelihood in \Cref{fig:ppd_all}. \textcolor{black}{Consequently, with the current data, one cannot distinguish a Gaussian from a log-normal LSS component. A larger number of host-identified FRBs is required for meaningful detection of this difference.} In particular, FRBs with higher redshift would be especially important to assess the effect due to the LSS field, as it becomes more dominant as the redshift increases. \review{Following the simplified forecasts done in \citet{hagstotz_2022_new}, we would expect that a couple of hundred FRBs can distinguish between these different models.}

\subsubsection{Variations in the Host component}
Now, we turn our focus on the host model of the DM, changing it from the fiducial log-normal to first a truncated Gaussian (t-Gaussian hereafter) and then to a Gamma distribution. The Gaussian is truncated at zero as the DM is positive by definition, hence, it is not Gaussian mathematically. For the t-Gaussian and Gamma hosts, the prior on $\sigma$ is uniform on [0, 500], which is broad enough to capture the expected long-tail behaviour. It also implies that the Gamma distribution can resemble either a log-normal or a t-Gaussian distribution depending on the data, as all of these distributions come from the exponential family. As before, the contours, $\chi^2$ and likelihood are shown in \Cref{fig:contour_plots} \Cref{fig:chi2} and \Cref{fig:ppd_all} respectively, all with $N_\mathrm{side}=512$. 

There are three columns for the three host models. The $\chi^2$-tests and their PTE values suggest that all models are good fits. The likelihoods of the DM for each model in \Cref{fig:ppd_all} also agree with the goodness of fit. As can be seen, all observed DMs are within the 25-75 percentile of the simulated DMs. In summary, the $\chi^2$-test does not prefer a particular model as the data currently lacks constraining power. The best-fit values in \Cref{tab:best_fit_all} for the host model show that we converge on the same behaviour. 

Looking at the best fit values, we can see that the statistical properties of the LSS field are rather sub-dominant with the current FRB sample and if a log-normal distribution is used for the host contribution. If we assume a (truncated) Gaussian host contribution, it cannot explain the high DM of FRBs at low redshifts, Hence, the model responds by artificially increasing the variance of the host contribution when the LSS is chosen to be Gaussian as well. In general, however, the constraints are all consistent with each other.

\begin{table}
\centering
\renewcommand{\arraystretch}{2.3}
\begin{tabular}{ccccc}

\textbf{Field} & \textbf{Host} & $\boldsymbol{\mathcal{A}}$ & $\boldsymbol{\mathrm{DM}_\mathrm{host}}\;[\mathrm{pc}\;\mathrm{cm}^{-3}]$ & $\boldsymbol{\sigma}$ \\
\hline\hline
\footnotesize
Log-$\mathrm{normal}^{\boldsymbol{*}}$ & log-$\mathrm{normal}^{\boldsymbol{*}}$  & $0.89_{-0.29}^{+0.31}$ & $223_{-95}^{+108}$ & $263_{-113}^{+91}$ \\
\hline
Log-normal & log-normal  & $0.93_{-0.22}^{+0.21}$ & $218_{-96}^{+86}$ & $250_{-86}^{+63}$ \\

Log-normal & t-Gaussian & $0.73_{-0.24}^{+0.22}$ & $340_{-170}^{+130}$ & $355_{-162}^{+73}$ \\

Log-normal & gamma & $0.94_{-0.25}^{+0.23}$ & $240_{-110}^{+130}$ & $260_{-140}^{+110}$ \\

Gaussian & log-normal & $0.79_{-0.24}^{+0.29}$ & $270_{-92}^{+102}$ & $205_{-67}^{+80}$ \\

Gaussian & t-Gaussian & $0.71_{-0.27}^{+0.35}$ & $220_{-150}^{+140}$ & $414_{-81}^{+58}$ \\

Gaussian & gamma & $0.82_{-0.24}^{+0.23}$ & $280_{-110}^{+120}$ & $297_{-76}^{+94}$ \\

\end{tabular}
\caption{Best-fit values with $1\sigma$ error bars for all the combinations of density fields and host models in our analysis are presented. The first row is our fiducial model with $N_\mathrm{side}=4096$ indicated by the $\boldsymbol{*}$ symbol. All the other values are calculated at a lower resolution of $N_\mathrm{side}=512$ for comparison. The corresponding contour plots are shown in \Cref{fig:contour_plots}. The $\sigma_\mathrm{LN}$ values for the log-normal host are converted to $\sigma$ for better comparison.}
\label{tab:best_fit_all}
\end{table}

\section{Conclusion}
\label{sec:conclusion}
In this paper, we have, for the first time, presented a simulation-based inference (SBI) analysis of the DM-$z$ relation, incorporating the appropriate statistical properties of the electron density field and the host contribution in forward simulations.

We introduced a novel set of simulations for DM observables, which can seamlessly incorporate any contribution to the DM along the line of sight. For the host contribution, we adopted a log-normal distribution as our fiducial setting, as it is widely accepted in the literature. However, we also implemented alternative functional forms of the host contributions, as this merely involves substituting a single function in the simulations. For the large-scale structure component, we utilised \texttt{GLASS} \citep{tessore_glass_2023}, which enabled us to simulate the electron density as either a log-normal or a Gaussian field with the correct correlations up to a given spatial resolution, provided by an input power spectrum of the three-dimensional electron field. The power spectrum was calculated using \texttt{HMCODE} \citep{mead_hydrodynamical_2020, tröster_2022_joint}, which was fitted to hydrodynamical simulations to jointly fit the matter and gas power spectra using a halo model approach. As output, we obtained concentric shells of the electron over-density field with a narrow width in redshift. FRBs were then placed in the electron density at their observed redshift and location from the real data (see below). After adding the stochastic host contribution, the line-of-sight integral was performed for each FRB and the Milky Way contribution was added. For the latter, we employed the standard method of using an electron model from prior literature \citep{cordes2002ne2001,yao_new_2017}, which is, in the spirit of our analysis, also fully flexible.
With this approach, we provided realistic simulated realisations of the DM given a cosmological model, which can be easily made more complex.

In the next step, we performed SBI on the described simulations. This inference method requires no explicit likelihood and works by training a NN to learn either the posterior distribution (non-amortised) or the joint distribution of data and parameters (amortised). In this context, the shape of the likelihood is unconstrained, which is why this approach is often referred to as likelihood-free inference (though it still implicitly requires a likelihood).
To demonstrate our pipeline, we applied it to 12 host-identified FRBs as a function of the cosmological amplitude of the DM-$z$ relation, $\mathcal{A}$ in \Cref{eq:DM_LSS_v2}, and two host contribution parameters. 
For our fiducial model, we used a log-normal electron density field as well as a log-normal host model. We inferred $\mathcal{A} = 0.89^{+0.31}_{-0.29}$, consistent with unity or the Planck cosmology. Similarly, the median and scale of the host distribution are $\mathrm{DM}_\mathrm{host}= 223^{+108}_{-95}\;\mathrm{pc \;cm}^{-3} \ \mathrm{and} \ \sigma_\mathrm{LN}= 0.76^{+0.37}_{-0.44}$. \review{The values for these three parameters are consistent with previous findings and we indeed found that no parameter (amplitude of the DM-$z$ relation, mean host contribution and its variance) is entirely dominated by its prior. This can be seen by the fact that the contours in \Cref{fig:contour_4096} are not entirely flat, even for $\sigma_\mathrm{LN}$. Following the discussion about the prior ranges in \Cref{tab:parameters}, this constraining power arises because very narrow ($\sigma_\mathrm{LN}\to 0$) and very broad ($\sigma_\mathrm{LN}>1.5$) are not supported by the data or by analytical models \citep[see also][]{2024arXiv241117682R}. Therefore, there is additional information in the data. This indicates that already 12 FRBs can inform us about the shape of the host contribution to some extent. }

The resulting posterior distributions were also assessed for consistency using standard coverage tests, specifically the TARP test. Our fiducial high-resolution case with a log-normal LSS and log-normal host component demonstrated perfect coverage, indicating that the learned posterior reproduces the coverage expected from random realisations from the simulator.

Furthermore, we assessed the quality of the fits using a Bayesian goodness-of-fit measure based on \citet{gelman1996posterior}, which was also employed in \citet{von_wietersheim-kramsta_kids-sbi_2024}. The $\chi^2$ was calculated from a number of data realisations generated from the simulations at the maximum a posteriori. This allows us to check whether the actual data is a plausible realisation of the likelihood. With the current data, we found all the models to be a good fit.

One of the added benefits of the SBI pipeline is that we can test the Gaussian likelihood assumption frequently used in a more traditional Bayesian approach. In that regard, we investigated the shape of the likelihood, finding the expected long-tailed distribution towards high DM values, which is necessary to explain the large DMs observed at low redshifts, such as those seen in FRB 20190520B. Another important check involved examining the evolution of the likelihood with redshift. We found that the mean of the probability distribution function increases with redshift, as expected. However, we also observed a reduction in its scatter. Although this behaviour may seem counterintuitive at first, it is supported by a straightforward analytical calculation, which shows that the width of the total distribution indeed reaches a minimum at redshift $z<1$ for the parameter values assumed in this analysis.

Lastly, we explored different modelling choices for both the LSS component and the host contribution. For the LSS component, we considered both Gaussian and log-normal distributions, while for the host contribution, we evaluated log-normal, truncated Gaussian and gamma distributions. We systematically tested all possible combinations of these models and found that they consistently provided similar constraints. To save computational time, we reduced the resolution of these simulations from \review{$N_\mathrm{side} = 4096$} to $N_\mathrm{side} = 512$. We found that the TARP test indicates that the learned posterior in these cases is slightly biased. \review{At lower $N_\mathrm{side}$, the variance of the LSS component is lower as small-scale effects are not accounted. Moreover, the interpolation of the $\texttt{HEALPix}$ maps also smooths out pixel values due to low pixel counts at lower resolution. Combined, these two results in smaller variance, which lead to a biased coverage. However, as all models are compared with the same resolution, the results from the comparisons still hold.} Importantly, none of the models showed any indication of being a poor fit to the data, which is mainly due to the low number of FRBs available with host identification at the moment.

In conclusion, the simulations and inference pipeline we have developed integrate the precise physical and statistical properties needed to accurately infer the DM-$z$ relation, free from general assumptions about the likelihood or posterior. This approach is highly adaptable and scales effectively with an increasing number of FRBs, thanks to the implemented data compression techniques. Moreover, our simulations provide a robust foundation for investigating systematic effects in cosmological studies involving FRBs, as these can be seamlessly incorporated at the map level. \review{It is, for example, easy to introduce any residual MW component which was not entirely removed by the models. Furthermore, introducing redshift and localisation errors and unlocalised FRBs into the analysis can be done easily in the simulation; thus,
looking ahead, a key direction will be to include more FRBs without known redshifts, enabling a joint fit of the DM-$z$ properties alongside the statistical properties of the DM. Implementing this in an analytic framework is intractable, making SBI an excellent alternative.}

The code for this work is publicly available via \href{https://github.com/koustav-konar/FastNeuralBurst}{https://github.com/koustav-konar/FastNeuralBurst}.

\section*{Acknowledgements}
SH was supported by the Excellence Cluster ORIGINS which is funded by the Deutsche Forschungsgemeinschaft (DFG, German Research Foundation) under Germany’s Excellence Strategy-EXC-2094-390783311.

\bibliographystyle{mnras}
\bibliography{MyLibrary_update,library2, library3}

\setcounter{figure}{0}
\renewcommand{\thefigure}{A.\arabic{figure}}
\renewcommand*{\theHfigure}{\thefigure}

\setcounter{table}{0}
\renewcommand{\thetable}{A.\arabic{table}}
\renewcommand*{\theHtable}{\thetable}

\newpage
\newpage
\appendix

\section{Effect of reducing $N_\mathrm{side}$}
\label{app:nsideeffect}
\begin{figure}
    \centering
    \includegraphics[width=0.45\textwidth]{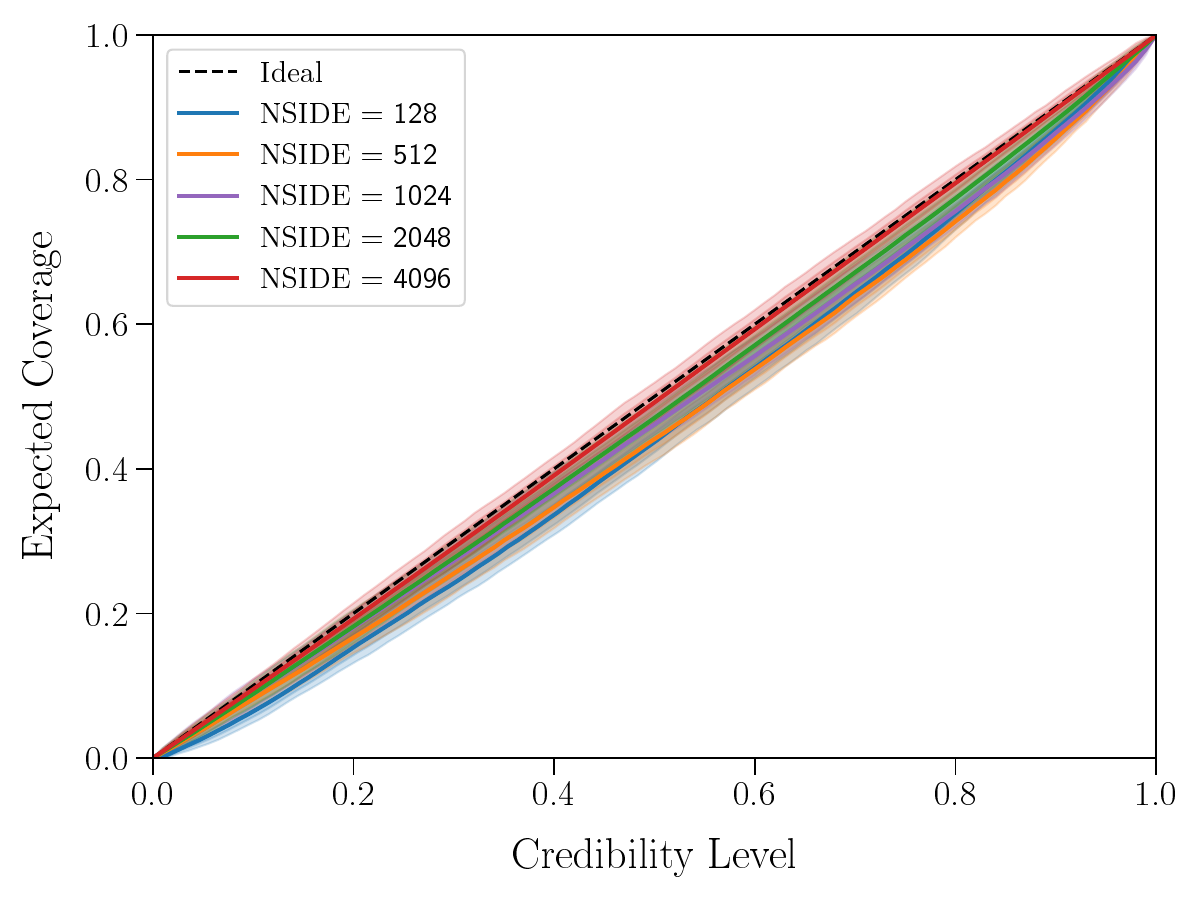}
    \caption{The multivariate TARP coverages for the fiducial model (using a log-normal distribution for both the field and the host) These are computed from 1000 forward simulations. One can see that an increased resolution of the maps increases the accuracy of the final posterior estimate. This behaviour is found in all cases studied in this work.}
    \label{fig:tarp}
\end{figure}
In \Cref{fig:tarp}, we show the effect of reducing the resolution of the \texttt{GLASS} forward simulation on the TARP coverage. This is necessary to establish the baseline for the comparison of the different models for the host and the LSS contribution, which have been run at a lower resolution. The main effect of decreasing $N_\mathrm{side}$ is that power on small scales gets washed out. In particular, the maximum multipole properly resolved is $\ell_\mathrm{max} = 3 N_\mathrm{side} - 1$ \review{\citep{tessore_glass_2023}.} However, loss of power already occurs earlier. In terms of effects on the inference process, this means that the variance introduced by the LSS component decreases and therefore, the scatter in the data will be larger than in the simulations. Therefore, the final constraints might be artificially tight. \review{Furthermore, inferring the DM contribution for a shell requires interpolation. Specifically, we take the bi-linear interpolation of the four neighbouring pixels. For a given $N_\mathrm{side}$, the total pixel count in a $\texttt{HEALPix}$ map is $N_\mathrm{pixel} = 12 N^2_\mathrm{side}$ \citep{gorski2005healpix}. Now, at a lower $N_\mathrm{side}$, the pixel count is lower and the interpolation effectively smooths out the expected fluctuation of $\mathrm{DM}_\mathrm{LSS}$, leading to an underestimate of error and bias in the TARP coverage.} \Cref{fig:tarp} demonstrates that the estimated posterior becomes slightly biased for decreasing $N_\mathrm{side}$ for the fiducial model. \review{We also note that the bias is a combined effect of the two scenarios mentioned above.} Since this will be true for all models and the fact that the high-resolution run is unbiased and exact, we can safely do a one-to-one comparison of the different models using a lower resolution.

\review{One potential solution to reduce the bias is to use weighted sampling of the $\texttt{HEALPix}$ maps to increase the LSS variance. However, we argue that it has little physical motivation. Therefore, as simulation-based inference (SBI) relies on the quality of simulations we use, it is advised to use a higher $N_\mathrm{side}$ for unbiased results. Alternatively, we can also reduce the threshold of the $1-\epsilon$ mass of the highest-probability region (HPR) of the approximate posterior in sequential training of TSNPE \citep[more details can be found in Section 6.10 in][]{deistler2022truncated}.}

\begin{figure*}
    \centering
    \includegraphics[width=0.45\linewidth]{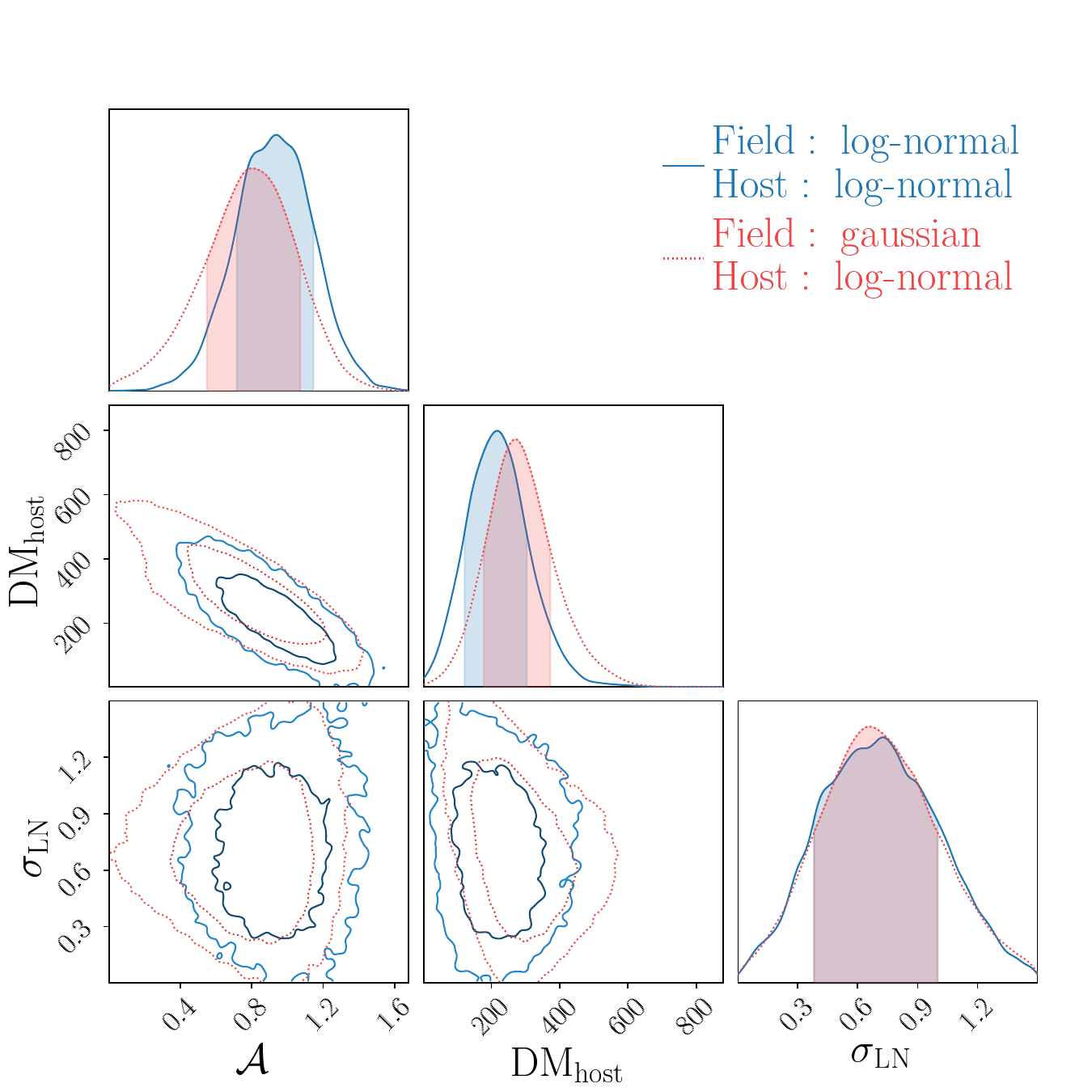}
\includegraphics[width=0.45\linewidth]{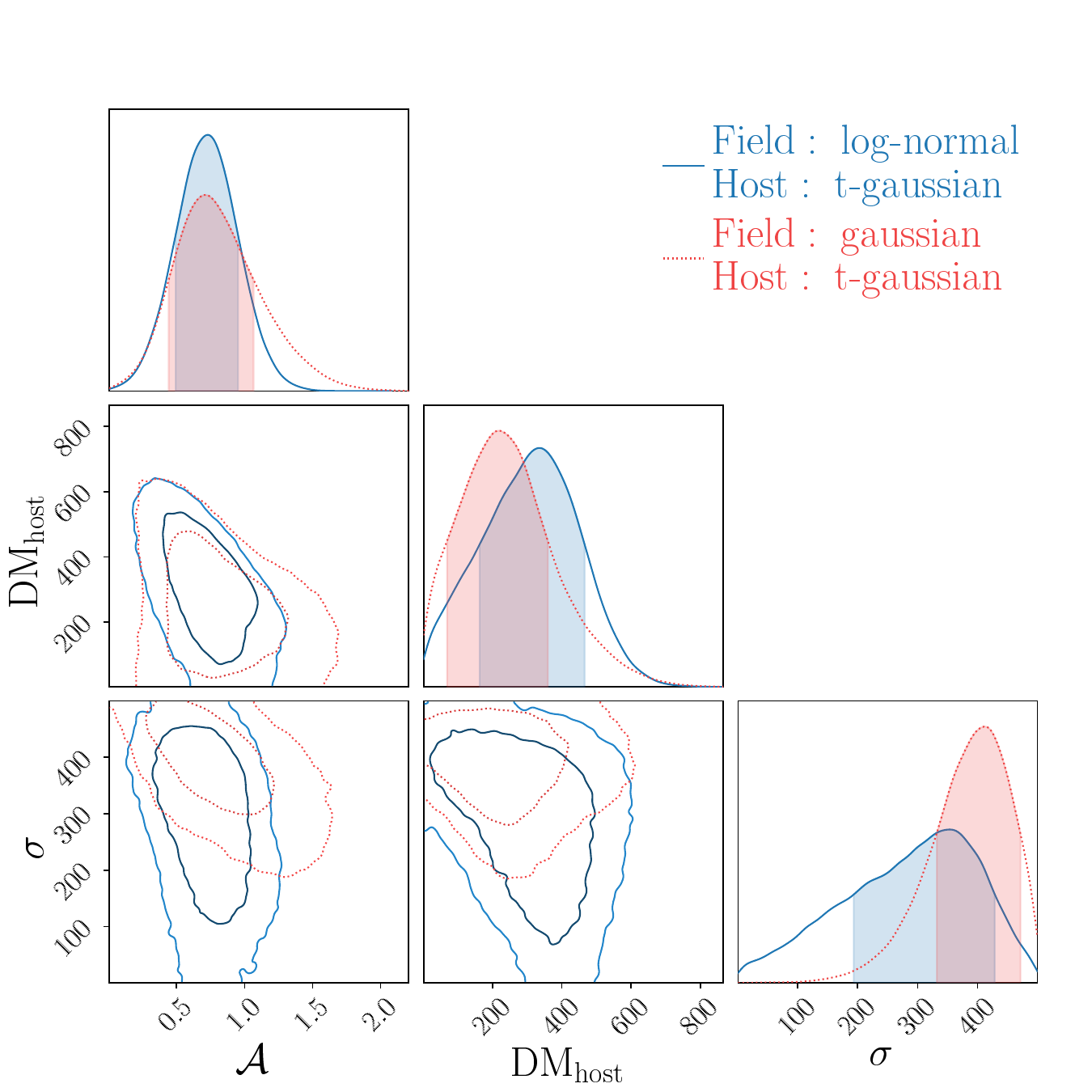}
\includegraphics[width=0.45\linewidth]{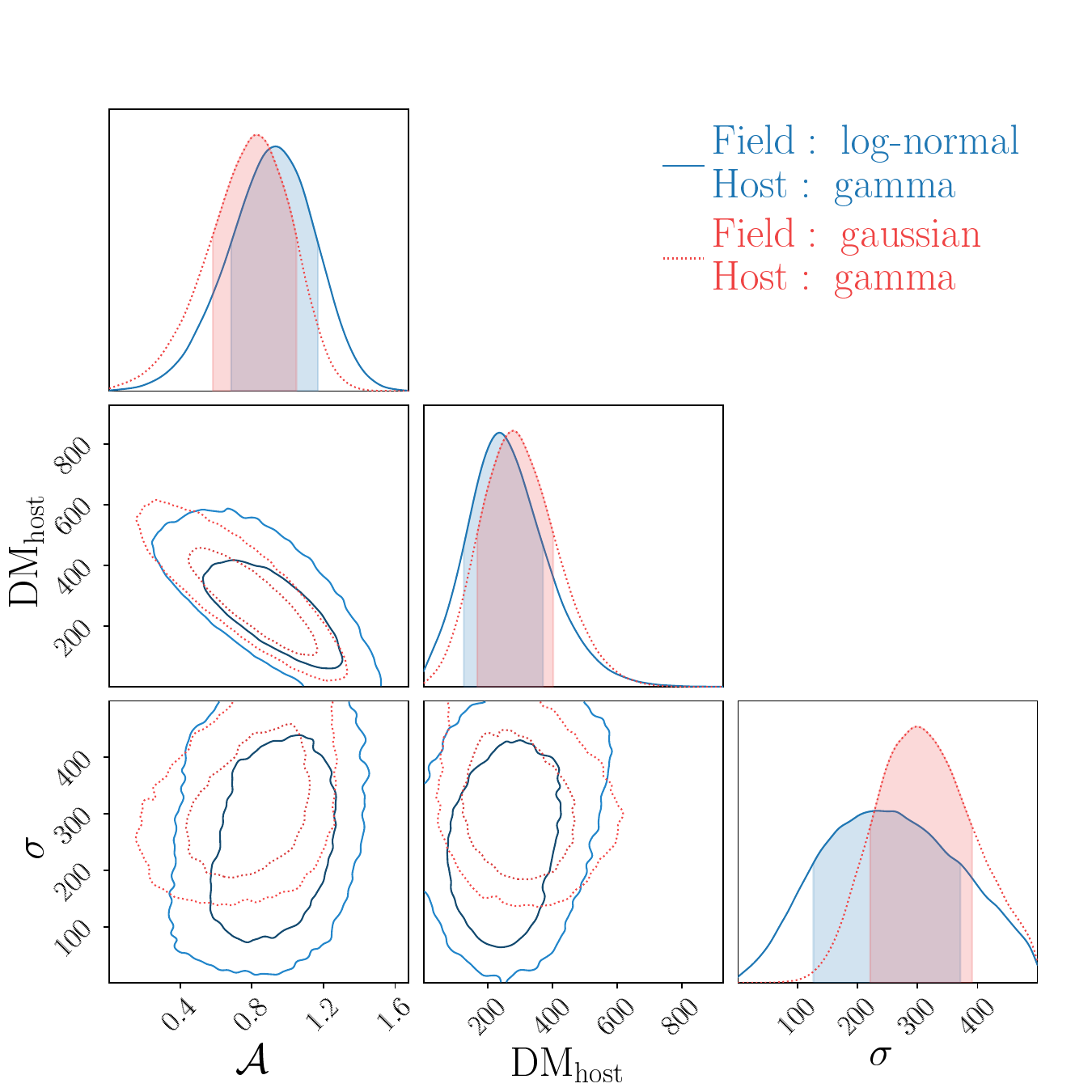}
    \caption{Marginalised contour plots for different combinations of the density field and host distribution. Each panel have the same host distribution and only the density field is varied, where the blue colour is for the log-normal field and red is for the Gaussian field. The host distribution for the top left, top right and bottom panels are log-normal, t-Gaussian and Gamma, respectively. The posteriors for all of these contours are evaluated with 1000 forward simulations in a 10-round TSNPE setup. The resolution of all the contours is  $N_\mathrm{side} = 512$ for comparison. The means with the $1\sigma$ errors are provided in \Cref{tab:best_fit_all}.}
    \label{fig:contour_plots}
\end{figure*}

\section{Results from different fits}
\begin{figure*}
    \centering
    \includegraphics[width=\linewidth]{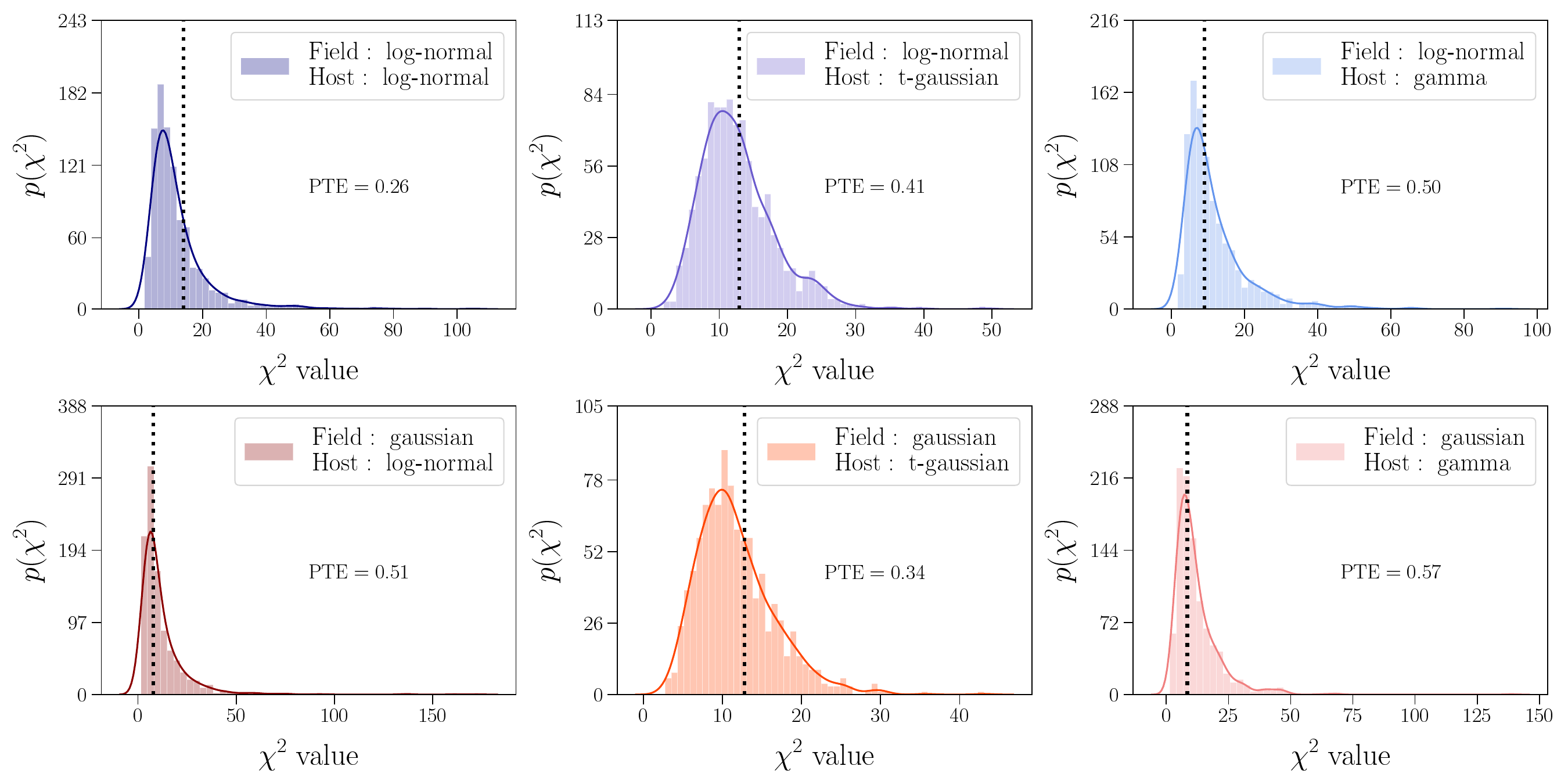}
    \caption{A comparison of the $\chi^2$ distribution for all the models we tested. The two rows represent the log-normal and Gaussian density fields respectively, while the three columns are for three different host distributions, i.e. log-normal,t-Gaussian and Gamma. The histograms are from 1000 forward simulations at best-fit values from \Cref{tab:best_fit_all} with $N_\mathrm{side} = 512$ for all cases. The corresponding probability mass beyond the observed $\chi^2$ (dotted line) is displayed as the PTE value.}
    \label{fig:chi2}
\end{figure*}

\begin{figure*}
    \centering
    \includegraphics[width=\linewidth]{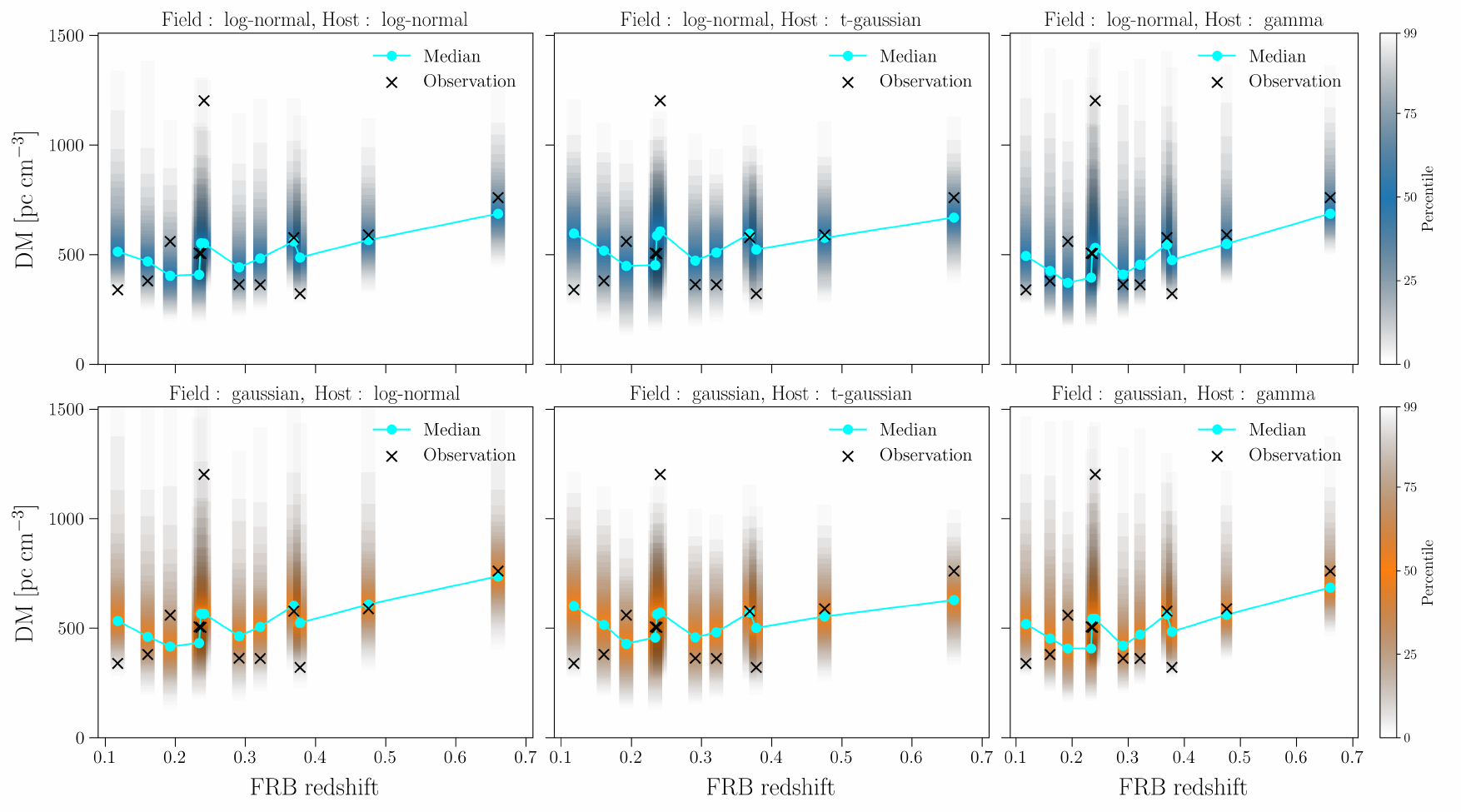}
    \caption{Likelihoods for all the models in our analysis. The two rows represent the log-normal and Gaussian density fields and the three columns are for the log-normal, t-Gaussian and Gamma host distribution. The simulated DMs for the individual FRBs are shown with a colour bar by mapping them to the percentile values. The cyan dots are the median and the black crosses are the observed DM.}
    \label{fig:ppd_all}
\end{figure*}

In this section, we present a comprehensive analysis of the different figures resulting from variations in both the host contribution and the large-scale structure (LSS) contribution. Specifically, we display the final contour plots \Cref{fig:contour_plots}, the goodness-of-fit tests in \Cref{fig:chi2} and the likelihoods in \Cref{fig:ppd_all}. 
Note that we use $N_\mathrm{side} = 512$ in this comparison to conserve computational resources. The reduced resolution underestimates the error from cosmic variance on individual measurements, leading to artificially tighter constraints. This issue has been addressed in our fiducial run, where a higher $N_\mathrm{side}$ was utilised to validate the pipeline, as described in the main text.
The key conclusion from these analyses is that all combinations of the model are statistically consistent with one another and provide a satisfactory fit to the data. This is primarily due to the relatively small number of FRBs considered in this study, as well as the conservative error estimates and the expansive posterior volume resulting from the use of a three-parameter model. While these effects limit the current discriminative power, they ensure robustness in our findings.

However, these tests will become increasingly critical as larger FRB samples become available in the future. The current coverage tests indicate that all posteriors exhibit a slight degree of bias when a Gaussian distribution is involved. 

Finally, for reference, we provide a list of the FRBs used in this analysis in \Cref{tab:frb_cat}.

\begin{table*}
\centering
\renewcommand{\arraystretch}{2.3}
\begin{tabular}{cccccccc}

\textbf{Name} & \textbf{Telescope} & \textbf{RA} & \textbf{DEC} & \textbf{Redshift} & $\boldsymbol{\mathrm{DM}_\mathrm{obs}}$ & $\boldsymbol{\mathrm{DM}_\mathrm{MW}}$ & \textbf{Reference} \\
\hline\hline

FRB 20190608A & ASKAP & 16:04.8	& -07:53:53.60 & 0.1178 & 339.50 & 204.75 & \citep{Chittidi_2021ApJ} \\

FRB 20200430A & ASKAP & 18:41.0 & +12:20:23.00 & 0.1608 & 380.10 & 112.21 & \citep{heintz_host_2020} \\

FRB 20150517A & Arecibo	& 32:01.0 & +33:07:56.00 & 0.19273 & 560.00 & 37.53 & \citep{chatterjee_direct_2017-1} \\

FRB 20191001A & ASKAP & 33:24.4 & -54:44:51.72 & 0.234 & 506.92 & 24.87 & \citep{heintz_host_2020} \\

FRB 20190714A & ASKAP & 15:55.1 & -13:01:14.52 & 0.2365 & 504.70 & 154.47 & \citep{heintz_host_2020} \\

FRB 20190520B & FAST & 02:04.3 & -11:17:17.32 & 0.241 & 1202.00 & 158.50 & \citep{niu_persistent_2022} \\

FRB 20190102C & ASKAP & 29:39.8	
 & -79:28:32.50 & 0.291 & 363.60 & 17.67 & \citep{bhandari_2020ApJ...895L..37B} \\

FRB 20180924B & ASKAP & 44:25.3 & -40:54:00.10 & 0.3212 & 362.40 & 34.43 & \citep{bannister_single_2019} \\

FRB 20200906A & ASKAP & 35:00.0 & -14:04:00.00 & 0.3688 & 577.80 & 91.27 & \citep{bhandari_characterizing_2022} \\

FRB 20190611B & ASKAP & 22:58.9 & -79:23:51.30 & 0.378 & 321.40 & 17.76 & \citep{heintz_host_2020} \\

FRB 20181112A & ASKAP & 49:23.6 & -52:58:15.39 & 0.4755 & 589.27 & 25.76 & \citep{prochaska_low_2019} \\

FRB 20190523A & DSA & 48:15.6 & +72:28:11.00 & 0.66 & 760.80 & 18.76 & \citep{ravi_fast_2019} \\
\end{tabular}
\caption{Table of localised FRBs used in our analysis with their respective parameters and references. The Milky Way (MW) contributions are calculated from the YMW16 model \footnote{\url{https://www.atnf.csiro.au/research/pulsar/ymw16/}} \citep{yao_new_2017} and have the same unit as $\mathrm{DM}_\mathrm{obs}$, i.e. $\mathrm{pc \; \mathrm{cm}^{-3}}$.}
\label{tab:frb_cat}
\end{table*}

\begin{figure}
    \centering
    \includegraphics[width=0.45\textwidth]{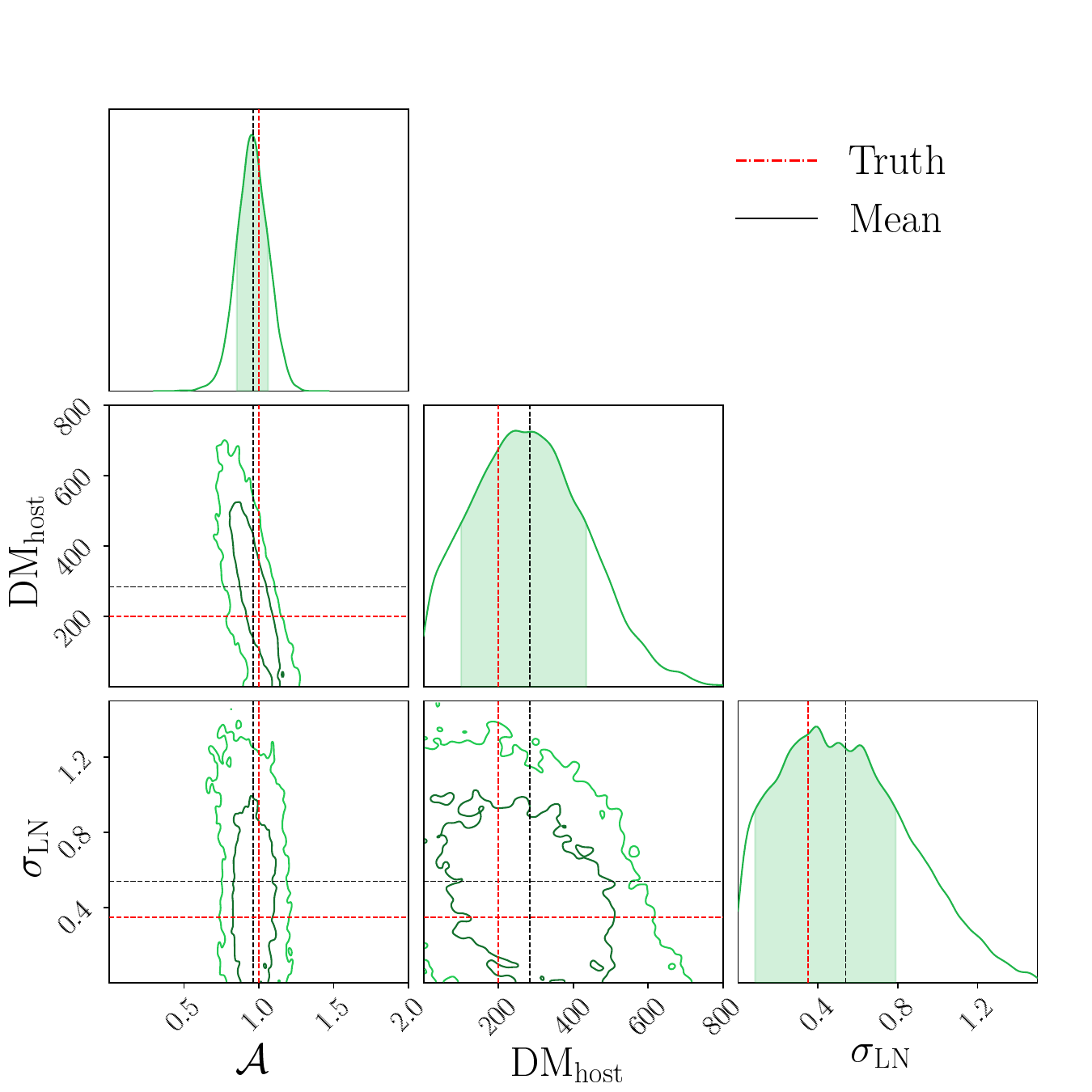}
    \caption{Marginalised contour plots with 100 mock FRB samples. The fiducial model with log-normal LSS field and log-normal host distribution is used. The truth (red line) is at (1, 200, 0.35) while the mean (black line) is at $\mathcal{A} = 0.96 \pm 0.11, \mathrm{DM}_\mathrm{host} = 284^{+100}_{- 200} \;\mathrm{pc \; \mathrm{cm}^{-3}}$ and $\sigma_\mathrm{LN} = 0.54^{+ 0.22}_{- 0.44}$.
}
    \label{fig:contour_with_mock}
\end{figure}

\begin{figure}
    \centering
    \includegraphics[width=0.45\textwidth]{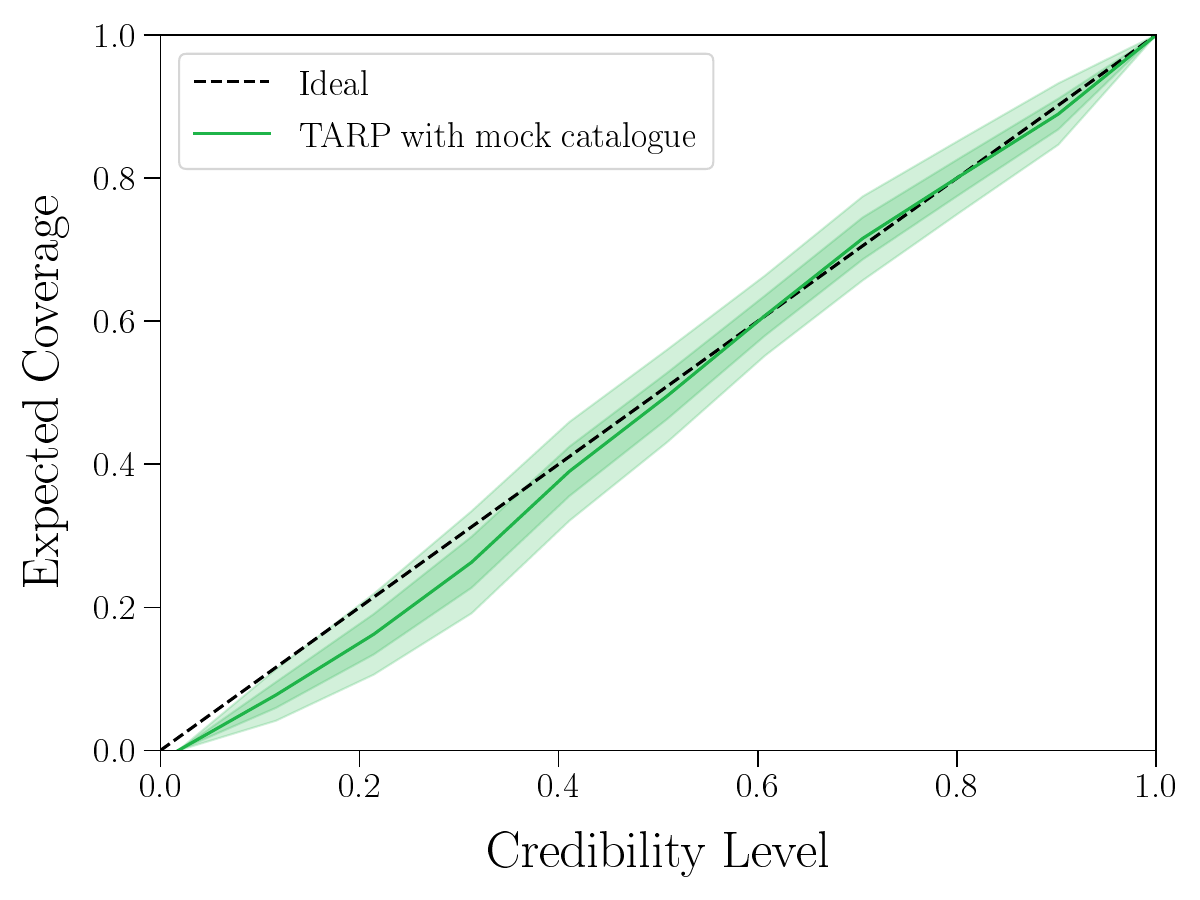}
    \caption{TARP coverage for the contour in \Cref{fig:contour_with_mock}. The simulations are with $N_\mathrm{side}=512$.}
    \label{fig:TARP_with_mock}
\end{figure}

\review{\section{Validation with mock catalogue}
To maintain end-to-end functionality and pipeline reliability, we ran the full analysis using a mock catalogue. The inputs in our pipeline are the right ascension (RA), declination (DEC), redshift ($z$), the Milky Way contribution ($\mathrm{DM}_\mathrm{MW}$) and observed dispersion measure ($\mathrm{DM}_\mathrm{obs}$) of the FRBs. For the test, we create a mock catalogue of 100 FRBs, where the RA and DEC are random samples drawn from a uniform distribution. Choosing the appropriate redshift distribution of FRBs is non-trivial as it closely relates to the formation mechanism, an area of ambiguity and active research. The distribution alters depending on the model, i.e. tracking the star formation history or compact mergers \citep[see e.g.][]{Munoz_2016_frb_redshift_dist, Wang_2020_frb_dist_mergers, reischke_probing_2021, James_2022_frb_sfr, Peng_2025_frb_redshift_distribution}. Hence, we use a more general distribution where the PDF is $z^2 \mathrm{exp}(-3z)$ and peaks at $z \approx 1$. Additionally, we consider a redshift range of $z = [0.1, 1]$, which mimics our analysis with the real data. The values of $\mathrm{DM}_\mathrm{MW}$ are taken from the YMW16 model \citep{yao_new_2017}. Finally, the simulator is run to create the mock dispersion ($\mathrm{DM}_\mathrm{obs}$). We use the fiducial model with a log-normal LSS field and log-normal host configuration where the parameters are set as $\mathcal{A} = 1, \mathrm{DM}_\mathrm{host} = 200$ and $\sigma_\mathrm{LN} = 0.35$. This constitutes the true values for the mocks, which are then used for the inference.

In \Cref{fig:contour_with_mock}, we show the contours from that analysis. The resolution parameter is $N_\mathrm{side}=512$ and all other parameters remain unaltered from the principal analysis. The mean values, indicated by the black line, are $\mathcal{A} = 0.96 \pm 0.11, \mathrm{DM}_\mathrm{host} = 284^{+100}_{- 200} \;\mathrm{pc \; \mathrm{cm}^{-3}}$ and $\sigma_\mathrm{LN} = 0.54^{+ 0.22}_{- 0.44}$. It demonstrates that the analysis can recover the true values (red line) effectively as the true values lie well within the $1\sigma$ range. A noteworthy aspect is that, despite having many more FRBs, the mock FRB sample is at higher redshifts than the real FRB sample. Therefore the host contribution is less dominant and therefore less constrainted than in the real analysis presented in this work. In contrast, however, the amplitude is much better constrained in the mock analysis for the same reason. 
Moreover, we check the coverage of the contours with the TARP coverage test and the result is shown in \Cref{fig:TARP_with_mock}. }

\end{document}